\documentclass[twocolumn,twocolappendix]{aastex631}

\usepackage{multirow}
\usepackage{xspace}
\usepackage{enumitem}
\usepackage[T1]{fontenc}
\usepackage{soul}

\newcommand{\gm}{\ensuremath{\gamma}}
\newcommand{\fermi}{\textit{Fermi}\xspace}
\newcommand{\swift}{\textit{Swift}\xspace}

\newcommand{\xmm}{\textit{XMM-Newton}\xspace}
\newcommand{\nustar}{\textit{NuSTAR}\xspace}
\newcommand{\bzq}{GB6~B1428+4217\xspace}

\newcommand{\msol}{\ensuremath{M_\odot}}
\newcommand{\gammasource}{J1429+420\xspace}
\newcommand{\rev}[1]{#1}

\begin{document}

\title{The most distant \gm-ray flare to date: a multiwavelength campaign on the $z=4.715$ blazar GB6~B1428+4217}

\email{dr.andrea.gokus@gmail.com, gokus@wustl.edu}
\author[0000-0002-5726-5216]{Andrea Gokus}
\affiliation{Department of Physics \& McDonnell Center for the Space Sciences, Washington University in St. Louis, One Brookings Drive, St. Louis, MO 63130, USA}

\author[0000-0002-1853-863X]{Manel Errando}
\affiliation{Department of Physics \& McDonnell Center for the Space Sciences, Washington University in St. Louis, One Brookings Drive, St. Louis, MO 63130, USA}

\author[0000-0002-3777-6182]{Ivan Agudo}
\affiliation{Instituto de Astrof\'isica de Andalucía (CSIC), Glorieta de la Astronomía s/n, 18008 Granada, Spain}

\author[0000-0002-8434-5692]{Markus B\"ottcher}
\affiliation{Centre for Space Research, North-West University, Potchefstroom 2520, South Africa}

\author[0000-0001-7112-9942]{Florian Eppel}
\affiliation{Julius-Maximilians-Universit\"at W\"urzburg, Fakult\"at f\"ur Physik und Astronomie, Institut f\"ur Theoretische Physik und Astrophysik, Lehrstuhl f\"ur Astronomie, Emil-Fischer-Str. 31, D-97074 W\"urzburg, Germany}
\affiliation{Max-Planck-Institut f\"ur Radioastronomie, Auf dem H\"ugel 69, D-53121 Bonn, Germany}

\author[0000-0002-4131-655X]{Juan Escudero Pedrosa}
\affiliation{Instituto de Astrof\'isica de Andalucía (CSIC), Glorieta de la Astronomía s/n, 18008 Granada, Spain}
\affiliation{Center for Astrophysics | Harvard \& Smithsonian, Cambridge, MA 02138, USA}

\author[0009-0009-7841-1065]{Jonas He{\ss}d\"orfer}
\affiliation{Julius-Maximilians-Universit\"at W\"urzburg, Fakult\"at f\"ur Physik und Astronomie, Institut f\"ur Theoretische Physik und Astrophysik, Lehrstuhl f\"ur Astronomie, Emil-Fischer-Str. 31, D-97074 W\"urzburg, Germany}

\author[0000-0001-6158-1708]{Svetlana Jorstad}
\affiliation{Institute for Astrophysical Research, Boston University, 725 Commonwealth Avenue, Boston, MA 02215, USA}
\affiliation{Saint Petersburg State University, 7/9 Universitetskaya nab., 199034 St. Petersburg, Russia}

\author[0000-0001-5606-6154]{Matthias Kadler}
\affiliation{Julius-Maximilians-Universit\"at W\"urzburg, Fakult\"at f\"ur Physik und Astronomie, Institut f\"ur Theoretische Physik und Astrophysik, Lehrstuhl f\"ur Astronomie, Emil-Fischer-Str. 31, D-97074 W\"urzburg, Germany}

\author[0000-0002-4184-9372]{Alex Kraus}
\affiliation{Max-Planck-Institut f\"ur Radioastronomie, Auf dem H\"ugel 69, D-53121 Bonn, Germany}

\author[0000-0002-3092-3506]{Michael Kreter}
\affiliation{Centre for Space Research, North-West University, Potchefstroom 2520, South Africa}

\author[0000-0001-6191-1244]{Felicia McBride}
\affiliation{Department of Physics and Astronomy, Bowdoin College, Brunswick, ME 04011, USA}

\author[0000-0001-9400-0922]{Daniel Morcuende}
\affiliation{Instituto de Astrof\'isica de Andalucía (CSIC), Glorieta de la Astronomía s/n, 18008 Granada, Spain}

\author[0000-0002-4241-5875]{Jorge Otero-Santos}
\affiliation{Instituto de Astrof\'isica de Andalucía (CSIC), Glorieta de la Astronomía s/n, 18008 Granada, Spain}
\affiliation{Istituto Nazionale di Fisica Nucleare, Sezione di Padova, 35131 Padova, Italy}

\author[0000-0003-2065-5410]{J\"orn Wilms}
\affiliation{Remeis Observatory \& Erlangen Centre for Astroparticle Physics, Universit\"at Erlangen-N\"urnberg, Sternwartstr.~7, 96049 Bamberg, Germany}

\begin{abstract}
In November 2023, the \textit{Fermi Large Area Telescope} detected a \gm-ray flare from the high-redshift blazar \bzq ($z=4.715$). We initiated a multi-wavelength follow-up campaign involving \swift, \nustar, the Sierra Nevada and Perkins Observatories, and the Effelsberg 100-m radio telescope. This source, also known as \rev{5BZQ\,J1430+4204}, has shown an anomalous soft X-ray spectrum in previous observations, including possible ionized absorption features or signatures of bulk Comptonization of thermal electrons\rev{, which are also detected during the flaring episode.} 
Simultaneous optical data revealed a polarization fraction of ${\sim}8$\% in the R band, confirming that synchrotron emission dominated over thermal emission from the accretion disk. The hard X-ray flux was enhanced during the flare. Modeling of the broadband spectral energy distribution suggests that the high-energy component is dominated by Compton scattering by external seed photons from the accretion disk. 
The origin of the flare is consistent with the injection of a hard-spectrum electron population in the emission region.
With a \gm-ray luminosity among the top 5\% of flaring events, \bzq exemplifies a prototypical MeV blazar. Its Compton-dominated SED and extreme luminosity are in line with expectations from the blazar sequence. High-redshift flares like this are critical for understanding jet physics in the early Universe and may improve detection prospects with future missions such as COSI.

\end{abstract}


\section{Introduction}\label{sec:intro}
Active galactic nuclei (AGN) are powered by accretion onto a supermassive black hole (SMBH). Their radiative output exceeds that of their host galaxy and makes them detectable across cosmological scales. 
The brightest AGN have generally been classified as quasars, with the most distant one being currently found at a redshift of $z=7.642$ \citep{wang2021}.
Interestingly, some of these AGN at high redshifts harbor SMBHs with masses exceeding $10^9$\msol\ \citep[e.g.,][]{ackermann2017, belladitta2020, burke2024, ighina2024}, indicating rapid black hole growth in the early Universe, within the first billion years after the Big Bang. 
Several theories exist for the evolution of these first SMBHs. They could have grown from light seeds with $M_{\mathrm{BH}}\leq10^2$\msol\ \citep[e.g.,][]{begelman2006, whalen2012}, heavy seeds with $M_{\mathrm{BH}}\geq10^4$\msol\  \citep[e.g.,][]{Bromm2003, wise2019}, or be the product of a series of hierarchical black hole mergers \citep[e.g., ][]{Volonteri2003}.
While `light' black holes are somewhat expected to be created early on as remnants of supernova explosions of the first stars \citep[e.g.,][]{madau2001}, their accretion rate would need to be consistently above the Eddington limit to reach a billion solar masses within a billion years. However, observational evidence shows that these extreme accretion states tend to be episodic \citep[e.g.,][]{Yu2002}.  
The existence of `heavy' seeds would relax the growth rate constraints \citep{bogdan2024}, but their proposed birth mechanisms still present theoretical challenges \citep[e.g.,][]{lodato2006,Inayoshi2020,cammelli2025}.

It is postulated that SMBHs in jetted AGN might grow faster, as powerful jets can enhance the accretion rate \citep[e.g.,][]{volonteri2011, ghisellini2013}. Conversely, simulations have shown that rapidly spinning black holes accreting at super-Eddington rates typically prompt the formation of powerful jets \citep[e.g.,][]{sadowski2014,mckinney2014}.
AGN that harbor a jet that we observe close to our line of sight are classified as blazars. 
Because blazar jets are pointing towards us, their emission appears particularly luminous and variable due to relativistic beaming \citep{urry1995}.
As such, blazars are ideal targets to study the conditions and growth of black holes in the early Universe.
Compact, flat-spectrum radio emission and detection of high-energy \gm-ray radiation are signatures indicating the presence of a blazar jet. In order to assess the jet power, observations in both the X-ray and \gm-ray range are crucial to constrain the shape of the high-energy component of a blazar's spectral energy distribution (SED). As their SED typically features two broad non-thermal components, the low-energy component originates from leptonic synchrotron emission, and in the case of high-$z$ blazars covers the radio up to the optical band. X-ray and \gm-ray observations cover the rise and fall of the inverse Compton hump, respectively. The peak of the component, which falls in the MeV range, can rarely be measured with current instruments due to the lack of sensitivity to faint AGN in the MeV band. 
In the high-energy regime, blazars have been detected well above $z>5$ at X-ray energies \citep[e.g.,][]{sbarrato2015, belladitta2020, medvedev2020, moretti2021, sbarrato2022, migliori2023, wolf2024, marcotulli2025}, with the most distant one recently being confirmed at $z=7$ by \citet{banados2024}. At \gm-ray energies, three sources have so far been detected above $z>4$ \citep{liao2018,kreter2020}. 
In the GeV band, detections are scarce due to the large distances as well as the source's redshift shifting the  emitted GeV radiation to the MeV band where the \fermi Large Area Telescope (LAT) is less sensitive. In addition, \gm-ray emission is being attenuated due to extragalactic background light above 10\,GeV \citep{dominguez2024}.

Despite these observational challenges, the detection of new high-redshift blazars is possible during flaring episodes. 
At X-ray energies, this behavior enabled the recently reported detection of a $z=6.19$ blazar in the hard X-ray band with \nustar \citep{marcotulli2025}.
In the \gm-ray band, \cite{kreter2020} showed that high-redshift sources that fall below the detection threshold of time-integrated searches of the \fermi-LAT data \citep{4fgl} can be significantly detected over a shorter time range of the order of 30\,days if the source is flaring. 
While nearby blazars show shorter typical flaring timescales \citep[days to weeks, e.g.,][]{Meyer2019}, 
the elapsed time in the rest frame of a high-$z$ source is related to the observer's time frame as $t_{\mathrm{rest}}=t_{\mathrm{obs}}/(1+z)$.
At $z\geq3$, a flare on a time scale of $t_{\mathrm{rest}}\sim7$\,days would correspond to $t_{\mathrm{obs}}\sim30$\,days.
Using this technique, and accounting for random fluctuations that could yield false positives,
\cite{kreter2020} reported detections of two $z>4$ blazars not previously listed in \fermi-LAT source catalogs.

Following this approach,
we monitor 80 blazars with $z\geq 3$ from the ROMA-BZCAT catalog \citep[5th edition;][]{bzcat_1stedition, bzcat_5thedition}, looking for a \gm-ray signal in the public real-time \fermi-LAT data using a 30-day integration window. 

In February 2022, our monitoring campaign led to the detection of an exceptional \gm-ray outburst from the blazar TXS\,1508+572 \citep[$z=4.31$, ][]{schneider2007}. The flaring activity continued for about six months and the \gm-ray luminosity peaked at about $5\times10^{49}$\,erg\,s$^{-1}$, placing it among the most luminous flares seen by \fermi-LAT \citep{Gokus2024}. 
Radio follow-up with the with the Very-Long Baseline Array (VLBA) and Effelsberg enabled a measurement of the kinematic speed of the jet \citep{benke2024}.

In this paper, we report the detection of a \gm-ray flare from an even more distant object, the blazar \bzq \citep[$z=4.715$;][]{Hook1998}.
To obtain the crucial quasi-simultaneous data at X-ray energies as well as data covering part of the synchrotron emission component, we launched a multiwavelength campaign using \swift, \nustar, the Perkins telescope, the T150 telescope at the Sierra Nevada observatory, and the 100-m Effelsberg radio telescope.

In Sect.~\ref{sec:flare_detection_with_lat}, we describe our analysis of the \fermi-LAT data and the source association of the \gm-ray detection with the high-$z$ blazar \bzq. We report on the multiwavelength data analysis in Sect.~\ref{sec:mwl_obs}, and present the results in Sect.~\ref{sec:results}. We discuss our findings and conclude with Sect.~\ref{sec:discussion}.
In this work, we assume a flat cosmology with $H_0=67.8\,\mathrm{km}\,\mathrm{s}^{-1}$, $\Omega_{\lambda}=0.692$, and $\Omega_M=0.308$ \citep{planck_collab2016}. 

\section{Detection of a gamma-ray flare from \bzq}\label{sec:flare_detection_with_lat}
\bzq, which is also known as B3\,1428+422 or 5BZQ\,J1430+4204, is the most distant known \gm-ray emitter as already reported by \citet{liao2018} and \citet{kreter2020}.  
Two distinct redshift values have been reported in the literature.
\bzq was first identified as a high-redshift radio source in 1998 by \cite{Hook1998}. Based on a spectrum taken with the 4.2\,m William Herschel telescope on La Palma, they derived a redshift of $z=4.715\pm0.010$ by averaging measurements of the Ly$\alpha$, \ion{Si}{4}/\ion{O}{4} and \ion{C}{4} emission lines. When referring to this work, the rounded up value of $z=4.72$ is often reported.
Later works rely on data from the Sloan Digital Sky Survey (SDSS). As part of an analysis for a large sample, \citet{richards2009} and \citet{sexton2022} reported redshifts of $z=4.665$ and $z=4.65$ based on photometric data and a Bayesian spectral analysis, respectively.
\citet{diana2022} measured the redshift of their sample of 19 high-redshift blazars using the \ion{C}{4} line and reported $z=4.71$ for \bzq.
For the sixteenth data release of the SDSS quasar catalog, different redshift values using different methods are reported, ranging from $z=4.64$ to $z=4.83$, but the most likely redshift after visual inspection is $z=4.705$ \citep{lyke2020}.
According to \citet{Hook1998}, the challenges in interpreting the spectrum arise primarily from an asymmetric Ly\,$\alpha$ line due to absorption and a significant velocity shift between the Ly\,$\alpha$ and \ion{C}{4} line. This makes the visual inspection of the spectrum that the authors conducted more likely to yield accurate values than the automated analyses done in other papers. In addition, the optical spectrum obtained by \citet{Hook1998} has better signal-to-noise ratio compared to the SDSS spectra discussed elsewhere in the literature. Consequently, we conclude that their reported $z=4.715$ is the most accurate redshift value for \bzq and we will use this value throughout the paper.
The resulting luminosity distance of \bzq is $\sim$44.6\,Gpc.

\bzq has been studied across the electromagnetic spectrum, in particular in the X-rays, where it is exceptionally bright.
A first observation with the High Resolution Imager \citep[HRI;][]{truemper1982,zombeck1995} onboard \textit{ROSAT} confirmed its quasar nature \citep{fabian1997}, with subsequent observations resolving intrinsic variability in the X-ray \citep{Boller2000} and radio bands \citep{fabian1999}.
Based on ASCA data revealing a hard X-ray spectrum, \bzq was proposed to belong to the class of MeV blazars \citep{fabian1998}.
\citet{Boller2000} found excess absorption using the Position Sensitive Proportional Counter \citep[PSPC;][]{truemper1982} instrument onboard \textit{ROSAT}, which was also detected in a \textit{BeppoSax} observation \citep{Fabian2001}, and suggested that it originates from a highly-ionized nuclear absorber with $\mathrm{N}_{\mathrm{H}}\sim10^{23}\,\mathrm{cm}^{-2}$.
Using the \textit{Chandra} observatory and the Very Large Array (VLA), \citet{Cheung2012} reported extended X-ray and radio emission from one of its jets.
In 2005, \bzq exhibited a radio flare that was not accompanied by an increase in X-rays \citep{worsley2006} nor the emergence of a new jet component \citep{veres2010}. However, long-term monitoring over 22\,years with VLBA revealed superluminal speed of its jet \citep{zhang2020}. The detection of \gm-ray emission associated with \bzq based on the observation of multiple subthreshold flares seen by \fermi-LAT was first reported by \citet{liao2018} and then confirmed by \citet{kreter2020}.

\subsection{Analysis of \fermi-LAT data}\label{sec:gamma_analysis}
Following the approach of \citet{kreter2020} we search for significant \gm-ray emission from high-redshift blazars in the real-time public LAT data using a 30-day integration window. In November 2023, our pipeline indicated the presence of a \gm-ray excess from the direction of \bzq in the MJD 60253.0--60283.0 time window.

After the inital trigger, we performed a dedicated analysis of the LAT data. To extract the data, we use ScienceTools Version 2.2.0 and \texttt{fermipy} version 1.3.1 \citep{fermipy} and select all events between 100\,MeV and 300\,GeV within a region of interest (ROI) of $15^\circ$ radius centered at the \gm-ray coordinates of the source to be analyzed. We select events of the \texttt{SOURCE} class and apply the  \texttt{DATA\_QUAL>0} and \texttt{LAT\_CONFIG==1} filters. In order to avoid contamination by photons originating from any Earth-limb effects, only events entering LAT with a maximum zenith angle of $90^{\circ}$ are considered. We use the post-launch instrument response functions \texttt{P8R3\_SOURCE\_V3}. 
We model the ROI considering all sources within $20^\circ$ of the ROI center that are listed in the 4FGL-DR4 catalog \citep{4fgl-dr4} and include isotropic (\texttt{iso\_P8R3\_SOURCE\_V3\_v1.txt}) and Galactic (\texttt{gll\_iem\_v07.fits}) diffuse \gm-ray emission. 
The method to model \fermi-LAT data is a maximum likelihood analysis that expresses the significance of a source in the model via the test statistic value $\mathrm{TS}=2\Delta\mathrm{log}(\mathcal{L})$. 
The test statistic thereby determines the difference in likelihood $\mathcal{L}$ between a model with and without a source \citep{mattox}. 
\begin{figure}
    \includegraphics[width=0.49\textwidth]{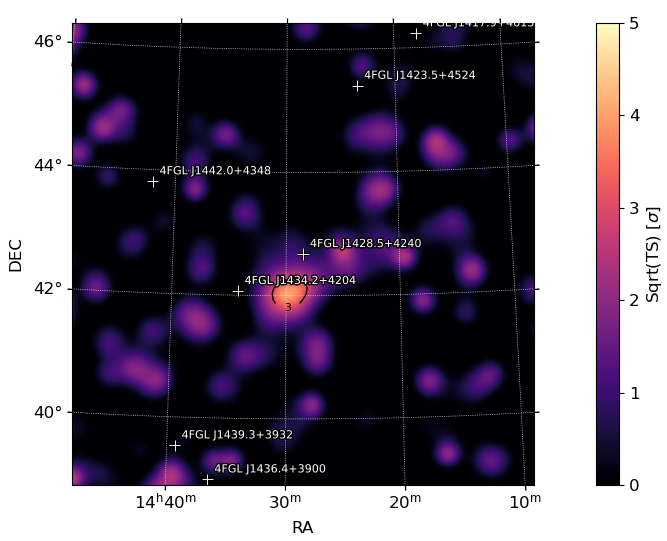}
    \caption{Cut-out of the resulting TS map for an analysis of the ROI without including a new \gm-ray source. The signals from known sources have been subtracted. An excess remains that is positionally coincident with the blazar \bzq.}
    \label{fig:tsmap_neighbours}
\end{figure}

\begin{figure}
    \centering
    \includegraphics[width=0.49\textwidth]{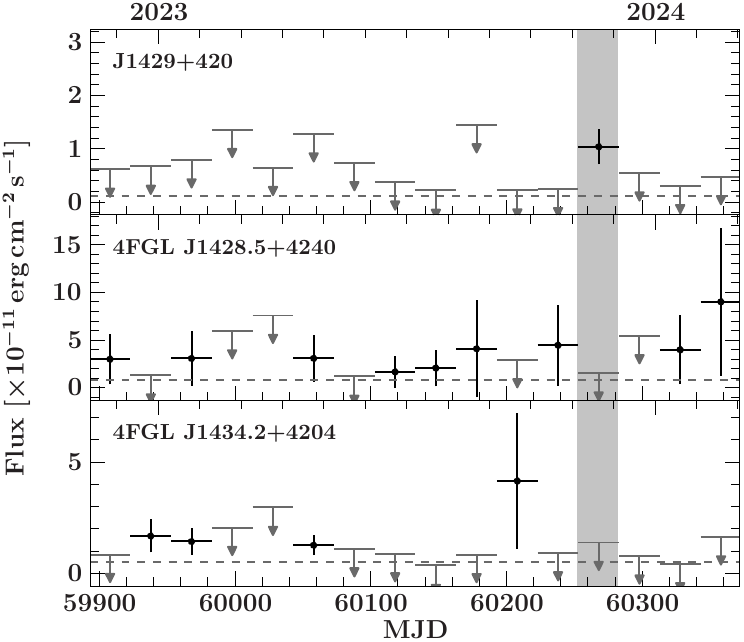}
    \caption{Gamma-ray light curves of \gammasource (\textit{top}), 4FGL~J1428.5+4240 (\textit{middle}), and 4FGL~J1434.2+4204 (\textit{bottom}). Each bin covers a time range of 30 days. Upper limits at the $2\sigma$ uncertainty are shown for bins that contain a low source detection significance of TS$<9$. The dashed horizontal line marks the average 4FGL flux of the two known \gm-ray sources, and the 15-year average flux derived for the new source \gammasource.}
    \label{fig:neighbor_lcs}
\end{figure}

The 4FGL catalog lists two \gm-ray sources, 4FGL~J1428.5+4240 and 4FGL~J1434.2+4204, within $1^\circ$ from \bzq. 
After optimizing our model for the ROI, we perform a likelihood fit, leaving the normalization and spectral parameters of 4FGL J1428.5+4240 and 4FGL J1434.2+4204 free. The normalization parameters for the Galactic and isotropic diffusion emission components as well as for point sources within 5$^{\circ}$ of the ROI are also left free.
Sources with $\mathrm{TS}< 1$ that are not within $3^{\circ}$ from the center are excluded from the model. 
The signals of 4FGL~J1428.5+4240 and 4FGL~J1434.2+4204 have a low significance with $\mathrm{TS}\sim 1$ and $\mathrm{TS}\sim 5$, respectively. The resulting TS map (see Fig.~\ref{fig:tsmap_neighbours}) shows an excess that cannot be satisfactorily explained by an enhanced $\gm$-ray flux from known 4FGL sources, indicating the presence of an additional \gm-ray source. 

We then proceed to redo the same analysis with a new source located at the position of the observed gamma-ray excess added to our model. 
To extract the signal from this new source, we fix the properties of all \gm-ray sources in the ROI after the initial optimization (using \texttt{optimize}) without the new \gm-ray source.
This is necessary to avoid fitting a model with too many free parameters that cannot be effectively constrained by the data.
In addition, for the closest neighboring sources, 4FGL~J1428.5+4240 and 4FGL~J1434.2+4204, we set the spectral parameters to their 4FGL value instead of using the optimized value. 
We choose this approach due to the spectral parameters of both sources agreeing with the 4FGL value and the possibility of low-energy photons originating from the \gm-ray excess being associated with the neighbors due to the large point spread function ($>1^\circ$) at $E<1$\,GeV. In particular, 4FGL~J1434.2+4204 is a flat spectrum radio quasar with a soft spectrum similar to what we would expect from a high-$z$ blazar, which could create confusion at the lowest energies.

Using the \textit{fermipy} \texttt{localize} function, we find that the new gamma-ray source, \gammasource, is located at $\alpha=217.41^{\circ}\pm0.08^{\circ}$, $\delta=+42.04^{\circ}\pm0.08^{\circ}$ 
with a localization uncertainty of $11.6’$ at 95\% confidence level. The null hypothesis \rev{(i.e.,} the new source is not present\rev{)} is rejected with a TS\,$=18.8$, corresponding to a significance of $\sim 4.3\sigma$.
The spectrum of the source is a power law with a photon index of $2.1\pm0.3$ and a flux of $(1.4\pm0.8)\times10^{-8}\,\mathrm{ph}\,\mathrm{cm}^{-2}\,\mathrm{s}^{-1}$, or $(1.3\pm0.6)\times10^{-11}\,\mathrm{erg}\,\mathrm{cm}^{-2}\,\mathrm{s}^{-1}$.
\rev{We acknowledge that the resulting significance is below the standard TS\,$\geq25$ used for reporting the detection of a new source. However, \bzq has already been established as a \gm-ray source based on recurring flare detections that individually appear at TS\,$>9$  \citep{liao2018,kreter2020}. 
The \gm-ray analysis reported here, together with the multiwavelength data collected contemporaneously, enables us to provide the best picture so far of the SED of this object.
}

In order to derive constraints for the \gm-ray emission before the detection of the flare, we perform a standard analysis using the 15-year LAT dataset (MJD 54683--60250) and add a point source at the best-fit coordinates derived for \gammasource.
In our analysis covering this time range, we detect an additional new \gm-ray source about $6^{\circ}$ from \gammasource, at $\alpha=215.672^{\circ}$ and
$\delta=48.005^{\circ}$, which is included in our analysis, and for which a follow-up analysis \rev{is under way for a future publication}.
The \gm-ray excess from \bzq in the 15-year LAT data has a significance of $\mathrm{TS}=15.3$, corresponding to ${\sim}3.9\sigma$. With a photon index of $3.1\pm0.3$, the spectrum is significantly steeper than during the flaring state. The derived flux is $(4.0\pm1.5)\times10^{-9}$\,ph\,cm$^{-2}$\,s$^{-1}$, or $(1.2\pm0.4)\times10^{-12}$\,erg\,cm$^{-2}$\,s$^{-1}$.

In addition, we compute 30-day binned light curves of \gammasource and both neighboring sources over a time range of $\sim15$ months. The light curves shown in Fig.~\ref{fig:neighbor_lcs} illustrate that the emergence of \gammasource is not coincident with increasing flux behavior in either of its neighboring sources. 

\subsection{Source association with the high-redshift blazar \bzq}\label{sec:source_assoc}

\begin{figure*}
    \centering
    \includegraphics[width=0.85\linewidth]{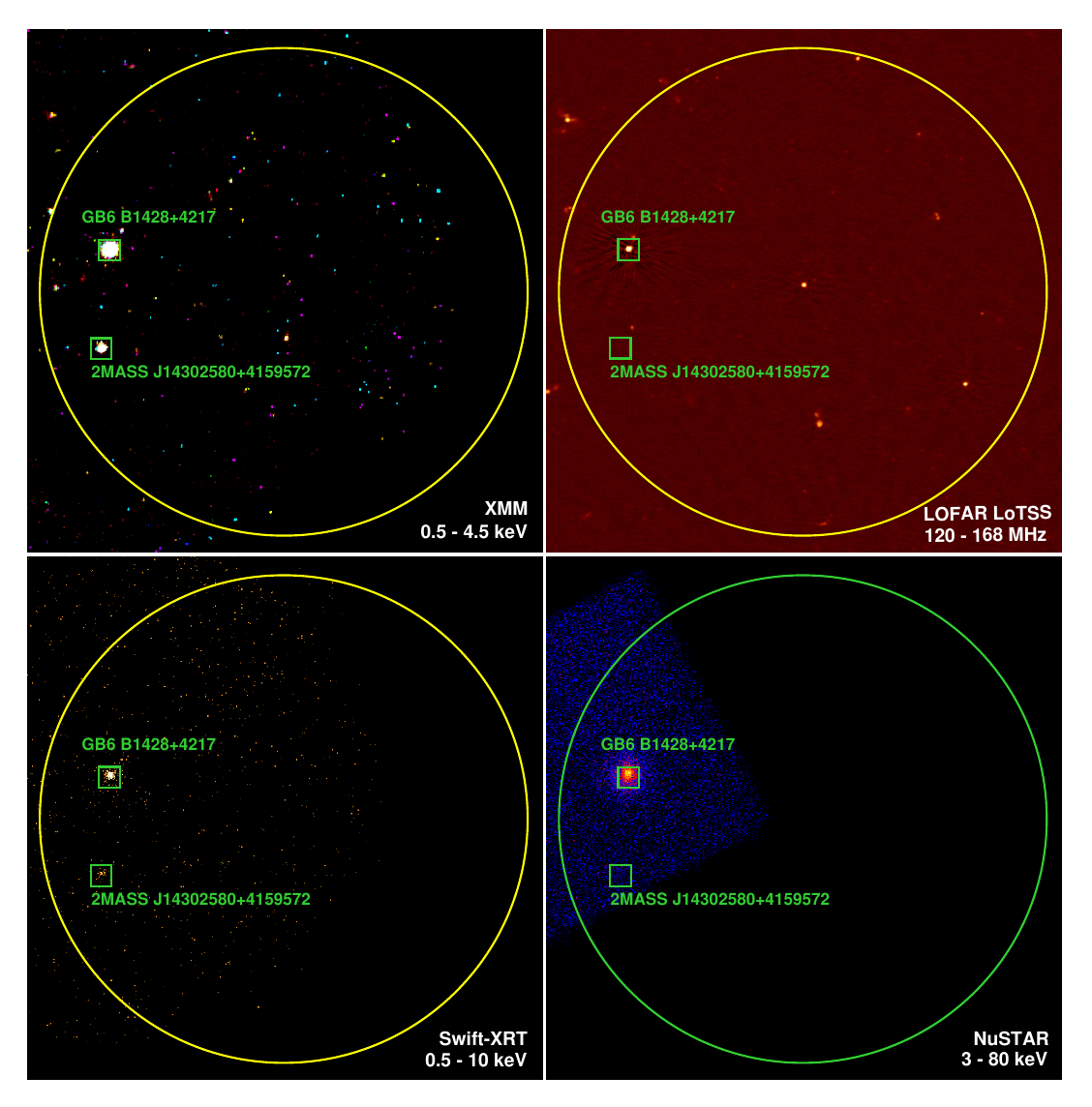}
    \caption{\textit{Upper panels:} Maps from \xmm \citep{xmm_serendipitous_source_catalog} and the LOFAR LoTSS \citep{lofar_lotts}. \textit{Lower panels:} \swift-XRT (ObsID: 00016413012) and \nustar (ObsID: 90901634002, FPMA) maps from observations taken as part of the multiwavelength campaign to follow-up the detection of \gammasource. Squares show \bzq and 2MASS~J14302580+4159572 centered on their radio and optical coordinates, respectively. The width of each map spans $\sim12'$, with the region of uncertainty being shown as a circle with a radius of $11.6'$.}
    \label{fig:ds9maps}
\end{figure*}

The large localization uncertainty of the new \gm-ray source of $11.6'$ requires a multiwavelength approach to determine the most likely counterpart of the detected \gm-ray source. We use publicly-available sky maps from \xmm \citep{xmm_serendipitous_source_catalog} and the LOFAR two-metre sky survey \citep[DR2;][]{lofar_lotts}, shown in Fig.~\ref{fig:ds9maps} (upper panels), to search for potential X-ray and radio counterparts within the 95\% uncertainty region of \gammasource.
In the \xmm X-ray image, which covers the complete uncertainty region, two bright sources are visible: the brighter one is the blazar \bzq ($\alpha= 217.598923^{\circ}$,$\delta=42.076803^{\circ}$), while the other prominent X-ray source is the quasar 2MASS~J14302580+4159572 ($\alpha=217.607442^{\circ}$, $\delta=41.999065^{\circ}$) located about $4.7'$ south of \bzq. 
In the radio image, more objects are visible, with the brightest again being \bzq. There is no radio counterpart for 2MASS~J14302580+4159572.
A radio source is close to the best-fit \gm-ray coordinates, however, the X-ray source with which it has been associated is too weak to be visible in the \xmm map. The \swift-XRT map (Fig.~\ref{fig:ds9maps}, bottom left panel) taken as part of our follow-up campaign also reveals \bzq and 2MASS~J14302580+4159572 as the only two prominent X-ray sources within the localization error. In the hard X-rays, about 20\% of the uncertainty area is covered, which includes both the positions of \bzq and 2MASS~J14302580+4159572. While \bzq appears as a bright source, only a slight excess is visible for 2MASS~J14302580+4159572. 
Hence, we conclude that the most likely counterpart for \gammasource is indeed \bzq. To probe the association with the Bayesian association method described in \citet{1fgl_cat}, we use our best fit coordinates, uncertainties, and ellipse position angle of 142.1$^{\circ}$ with the prior probability value for the BZCAT (0.308). Our calculation yields a probability of 0.87 for \bzq being the counterpart of \gammasource.

\section{Multiwavelength observations and data analysis}\label{sec:mwl_obs}
\subsection{X-ray data}
We model all X-ray spectra with the Interactive Spectral Interpretation System \citep[ISIS, Version 1.6.2-51][]{isis} using \texttt{vern} cross sections \citep{vern} and the \texttt{wilm} abundances \citep{wilms_tbabs} for absorption by the interstellar medium. 
The Galactic foreground absorption for \bzq is $N_\mathrm{H}=9.29\times10^{19}$\,cm$^{-2}$ according to the HI 4$\pi$ survey \citep[HI4PI; ][]{HI4PI}.
All uncertainties of the best-fit parameters are given at the $1\sigma$ confidence level.

\subsubsection{\swift-XRT}\label{sec:xrt_extraction}
We extract all available \swift-XRT data in order to 
characterize the X-ray flux variability from \bzq.
This blazar has been observed by \swift on singular instances in 2008, 2013, 2014, and 2023, while in 2021 and 2024 denser monitoring is available. The latest monitoring data set is connected to our follow-up campaign after the detection of the \gm-ray flare.
We extract spectra and corresponding response files using HEAsoft \citep[V6.33.2;][]{heasoft} and the XRTDAS\footnote{\url{https://swift.gsfc.nasa.gov/analysis/xrt_swguide_v1_2.pdf}} pipeline (v 3.7.0) with CALDB version 20240522.
All observations have been performed in Photon Counting mode. As source extraction region, we chose a circular region of $40''$ centered on the source coordinates ( $\alpha=217.59827^{\circ}$, $\delta=42.07589^{\circ}$), while the background region is chosen as an annulus with an inner radius of $50''$ and an outer radius of $150''$ centered on the same coordinates.

\subsubsection{\nustar}
We obtained a target-of-opportunity observation with \nustar on 2023 Dec 09 (ObsID: 90901634002) for a total exposure time of 58\,ks. For the data extraction, we used the standard procedures with \texttt{NUSTARDAS}\footnote{\url{https://heasarc.gsfc.nasa.gov/docs/nustar/analysis/nustar_swguide.pdf}} (v~2.1.2) and CALDB version 20231205.
Using \texttt{nuproducts}, we extract response files and spectra of the source from a circular region with a radius of $60''$ and centered on $\alpha=217.5997^{\circ}$ and $\delta=42.0710^{\circ}$ for both Focal Plane Modules A and B (FPMA, FPMB).
For the extraction of the background spectra, we chose an annular region around the source coordinates with inner and outer radii of $60''$ and $120''$, respectively. 

In addition, we extract spectra from an earlier observation of \bzq that was taken in July 2014 (ObsID: 60001103002) using the same procedure and tools. For both FPMs, the extraction region of the source spectra is centered on slightly different coordinates (RA\,=\,217.5995$^{\circ}$ and Decl.\,=\,42.0760$^{\circ}$) but the same radius for the region is chosen. The background spectra are extracted from circular regions in a source-free region with a radius of 120''.
The exposure of this observation is 49\,ks.

\subsection{Optical}
In December 2023, we monitored \bzq with the 1.5-m telescope at the Sierra Nevada Observatory and obtained optical photometry in the R and I band. In addition, we measured the optical polarization of the source in the R band with the 1.83\,m Perkins telescope.
We note that due to the high redshift of this blazar ($z=4.715$), optical data with wavelengths shorter than 694.8\,nm / $4.315\times10^{14}$\,Hz are affected by absorption of the Lyman-alpha forest ($\lambda_{\mathrm{rest}}=121.57$\,nm).

\subsubsection{Sierra Nevada Observatory}
\begin{table}
\centering
\caption{Optical data taken with the Sierra Nevada Observatory (SNO) and the Perkins telescope (PTO).}\label{tab:optical_data}
\begin{tabular}{lcccc}
\hline\hline
Date     & MJD     & \multicolumn{2}{c}{Magnitude}       & Observatory \\
 & & R band {[}mag{]} & I band {[}mag{]} &    \\ \hline
2023-12-12       & 60290.2 & $20.50\pm0.30$     & $19.21\pm0.23$   & SNO \\
2023-12-14       & 60292.2 & $20.53\pm0.25$   & $19.44\pm0.20$   & SNO \\
2023-12-16       & 60294.2 & $20.58\pm0.21$   & $19.25\pm0.13$   & SNO \\
2023-12-19       & 60297.2 & $20.67\pm0.12$   & $19.54\pm0.24$   & SNO \\
2023-12-20       & 60298.2 & $20.25\pm0.29$   & $19.50\pm0.40$     & SNO \\
2023-12-22       & 60300.2 & $20.42\pm0.09$   & $19.64\pm0.15$   & SNO\\
2024-01-10       & 60319.5 & $20.60\pm0.07$   & -& PTO\\ \hline
\end{tabular}
\end{table}

Optical photometric observations of \bzq were also performed from Sierra Nevada Observatory (Granada, Spain). These observations were carried out with the 1.5-m T150 telescope in the optical R and I bands between 2023 December 12 and 22, within the blazar optical monitoring program TOP-MAPCAT \citep{agudo2012}. These data were reduced and analyzed with the automatic photo-polarimetric analysis pipeline IOP4 \citep[see][]{escudero2024a, escudero2024b}.
We list our results in Table~\ref{tab:optical_data}.
For 2023 December 22, we obtained three observations taken about 20 to 30\,minutes apart each. Since within uncertainties the measured magnitudes agreed with one another, we list the average magnitude calculated using the inverse variance weighting method in our results table.
We do not find noticeable variation of the source during the monitoring period in December 2023 and it also does not appear brighter than an archival observation in the R band listed in 5BZCAT\citep{bzcat_1stedition}.

\subsubsection{Perkins telescope}
We performed photometric and polarimetric observations of the quasar \bzq on 2024 January 10. The observations were carried out in R band ($\lambda_{\text eff}$\,=635~nm) at the 1.83\,m Perkins telescope (PTO, Flagstaff, AZ, USA) using the PRISM camera\footnote{\url{https://www.bu.edu/prism/}} equipped with a polarimeter with a rotating half-wave plate. 
The polarimetric observation involved three series of Stokes parameters $Q$ and $U$ measurements. Each series consisted of four measurements at instrumental position angles (P.A.s) $0^{\circ}$, $90^{\circ}$, $45^{\circ}$, and $135^{\circ}$, with a 300\,s exposure at each P.A. 
Each Stokes parameter was averaged over series, with the scatter across the series used to estimate an uncertainty.  
The PRISM camera has a field of view of $14'\times14'$. This allows us to use field stars to perform both interstellar and instrumental polarization corrections of the Stokes parameters. For absolute calibration of the polarization P.A., we used polarized stars from \cite{Schmidt1992}. 
After all corrections we have obtained a final degree of polarization 
$\mathrm{P}=(8.6\pm3.0)$\% and electric-vector P.A. $\chi=(-49.5\pm9.8)^{\circ}$ for the quasar \bzq.

\begin{figure}
    \centering
    \includegraphics[width=0.5\textwidth]{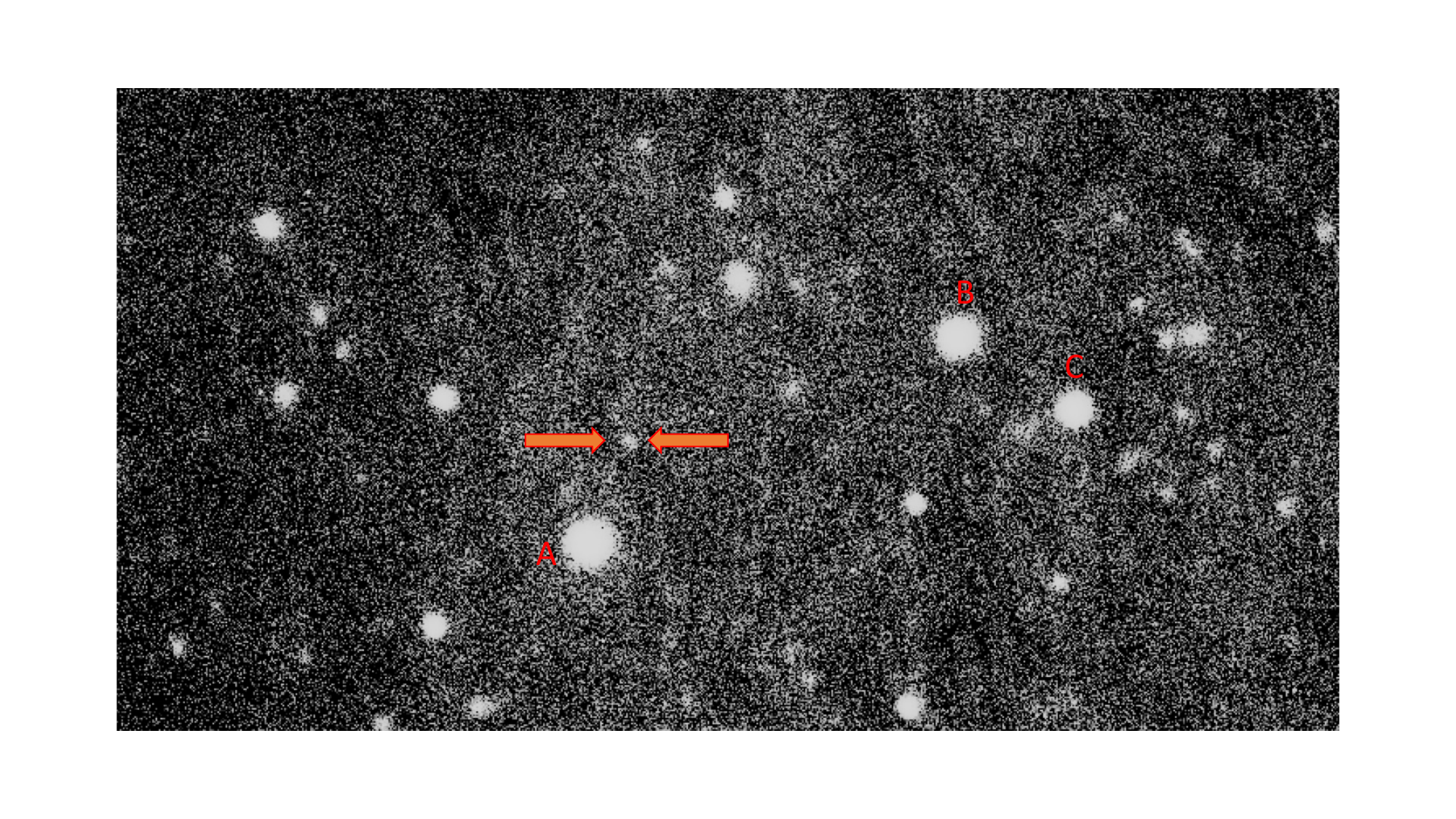}
    \caption{Field of \bzq in R band in polarized light, labels show the comparison stars A, B, and C.}
    \label{fig:comparison_star_field}
\end{figure}

We have determined magnitudes in R band of three comparison stars, A, B, and C (see Fig. \ref{fig:comparison_star_field}), in the field of \bzq using differential photometry with comparison stars 1, 2, 3, and 4 in the field of blazar 1ES~1426+428 with known magnitudes \citep{Smith1991}, which is  close to the field of \bzq. 1ES~1426+428 was observed with three exposures of 60\,s each before and after observations of \bzq. \bzq was observed with three exposures of 300\,s. As a result of these observations, we have determined the R band magnitudes of the stars in the field of \bzq as follows: A=$15.121\pm0.005$, B=$15.721\pm0.006$, C= $16.674\pm0.004$. These stars were used to estimate the magnitude of the quasar in the R band on 10 January 2024 to $20.60\pm0.07$. 

\subsection{Radio data}

\begin{table}
\centering
\caption{Radio measurements taken with the 100-m radio telescope Effelsberg.}\label{tab:radio_data}
\centering
\begin{tabular}{lccc}
\hline\hline
Date   & MJD  & Frequency & Flux density \\ 
 &   & [GHz] & [mJy] \\ \hline
\multirow{4}{*}{2023-12-15} & \multirow{4}{*}{60293.5} & 4.85  & $111\pm11$  \\
   &    & 6.55   & $105\pm2$   \\
   &    & 15   & $176\pm3$    \\
  &     & 21.4    & $163\pm4$ \\ \hline
\multirow{3}{*}{2024-01-14}  & \multirow{3}{*}{60323.5} & 2.35  & $110\pm1$  \\
    &    & 4.85   & $109\pm1$  \\
    &    & 10.45   & $144\pm2$   \\ \hline
\end{tabular}
\end{table}

We observed \bzq with the Effelsberg 100-m telescope on 15 December 2023
and 14 January 2024. These observations covered a range of frequencies from
${\sim}$2\,GHz to $\sim$25\,GHz. We performed cross-scans over the position of the
point-like source in azimuth and elevation. The data are then averaged,
undergo a quality-check by a semi-automatic pipeline, and are corrected for
pointing offsets, atmospheric opacity and elevation-dependent gain
errors. Our calibration source is the standard calibrator 3C~286. 
The details of the observation and data reduction process are
described in \citet{eppel2024}. While we followed the same strategy, we added
observations at lower frequencies. 
We list the fluxes from both observing epochs in Table~\ref{tab:radio_data}.
During both observations, we find flux densities of $\sim$110\,mJy at lower frequencies (2--5\,GHz), while the flux densities increase at higher frequencies up to 176\,mJy at 15\,GHz, resulting in a slightly inverted spectrum.
Compared to archival data presented in \citet{worsley2006}, the measured radio fluxes are comparable to a non-flaring state in the radio band.

\section{Results}\label{sec:results}

\subsection{Gamma-ray spectra}\label{sec:gammaray_results}

\begin{table}
\centering
\caption{Gamma-ray best fit results.}\label{tab:gamma-ray_fit_results}
\centering
\begin{tabular}{lccc}
\hline
\multirow{2}{*}{MJD} & Flux$_{0.1-300\,\mathrm{GeV}}$    & \multirow{2}{*}{$\Gamma$} & L$_{0.1-300\,\mathrm{GeV}}$   \\
   & {[}$10^{-8}$\,ph\,cm$^{-2}$\,s$^{-1}${]} &      & {[}$10^{48}$\,erg\,s$^{-1}${]} \\ \hline
60253 -- 60283  & $1.4\pm0.8$  & $2.1\pm0.3$  & $3.9\pm3.8$  \\
54683 -- 60250  & $0.40\pm0.15$  & $3.1\pm0.3$  & $2.0\pm0.8$  \\ \hline
\end{tabular}
\end{table}
Even though we average over $\sim$15\,years of LAT data, the signal determined for our long-term spectrum and the derived flux and photon index may be contaminated by flaring episodes.
Our best fit results for the \gm-ray spectra are listed in Table~\ref{tab:gamma-ray_fit_results}.
The flux increase during the flaring epoch in 2023 in comparison to the flux derived from the long term measurement is a factor of ${\sim}3.5$, however due to the large uncertainties, a 22\% chance remains that those fluxes are in agreement with one another.
The blazar seems to exhibit a harder-when-brighter behavior found in the majority of FSRQ. The difference in photon index is of the order of $2.4\sigma$, that is, only a small probability of 1.8\% that the \gm-ray photon index of \bzq did not change between the long-term and the flare data sets. 
The measured luminosity during the flare in 2023 is poorly constrained, and does not get close to the record \gm-ray luminosities measured for other high-$z$ blazars during flaring epochs \citep[e.g., TXS\,1508+572 and PKS\,0537$-$286,][respectively]{Gokus2024,sahakyan2020}.

\subsection{X-ray spectra}\label{sec:xrayspectra}
Using data taken simultaneously with \swift-XRT and \nustar, we search for the presence of a spectral break and  test for the previously reported strong soft X-ray absorption \citep{Boller2000,Fabian2001,Worsley2004,worsley2006}.
For all modeling reported in this section, we rebin each spectrum using the optimal binning algorithm by \citet{optimalbinning}.

\begin{table*}
\centering
\caption{Best-fit results for an absorbed power law model \rev{(\texttt{tbabs*ztbabs*cflux(powerlaw)})} applied to two contemporaneous data sets of \nustar and \swift-XRT spectra taken in 2014 and 2023, respectively. $N_{\mathrm{H}}$ \rev{lists the source-intrinsic cold absorption while} $N_{\mathrm{H, Gal.}}$ \rev{is fixed to the Galactic foreground value.}}\label{tab:xray_fitresults_nh_constraint}
\centering
\begin{tabular}{lccccccccc}
\hline
Date   & MJD & Inst. & ObsID   & Exposure & Photon Index   & Flux$_{0.3-80\,\mathrm{keV}}$ & $N_{\mathrm{H}}$ & $\chi^2$/d.o.f.  \\ 
 & &  &  & [ks] &   & {[}erg cm$^{-2}$ s$^{-1}${]} & {[}cm$^{-2}${]}&  \\ \hline
2014-07-14 & 56852.0 & N & 60001103002 & 49.2  & \multirow{3}{*}{$1.67\pm0.04$} & \multirow{3}{*}{$(5.1\pm0.2)\times10^{-12}$}   & \multirow{3}{*}{$(\rev{6.5}^{+4.4}_{-2.9})\times10^{\rev{22}}$} & \multirow{3}{*}{\rev{451.5} / 381} \\
2014-07-02 & 56840.6 & S & 00080752001 & 2.2   &    & & &  \\
2014-07-13 & 56851.0 & S & 00080752002 & 7.4   &    & & &  \\ \hline
2023-12-09 & 60287.0   & N & 90901634002 & 58.7  & \multirow{8}{*}{$1.506\pm+0.025$} & \multirow{8}{*}{$(1.03\pm0.03)\times10^{-11}$} & \multirow{8}{*}{$(\rev{5.4}^{+1.4}_{-1.2})\times10^{\rev{22}}$} & \multirow{8}{*}{\rev{517.9 / 607}} \\
2023-12-06 & 60284.7 & S & 00016413001 & 3.0  &    & & &  \\
2023-12-07 & 60285.6 & S & 00016413002 & 2.0  &    & & &  \\
2023-12-08 & 60286.5 & S & 00016413003 & 2.8  &    & & &  \\
2023-12-09 & 60287.2 & S & 00016413004 & 2.6  &    & & &  \\
2023-12-10 & 60288.5 & S & 00016413005 & 2.8  &    & & &  \\
2023-12-11 & 60289.1 & S & 00016413006 & 2.8  &    & & &  \\
2023-12-12 & 60290.1 & S & 00016413007 & 2.4  &    & & &  \\ \hline
\end{tabular}
\end{table*}

\begin{figure}
    \centering
    \includegraphics[width=0.95\linewidth]{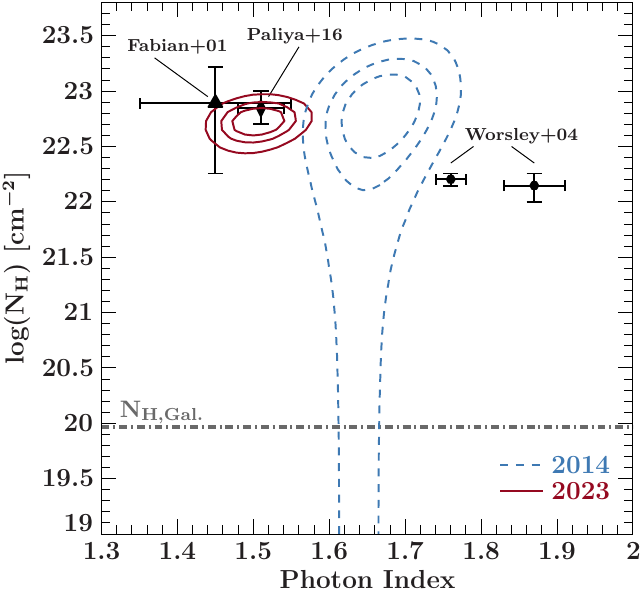}
    \caption{Confidence maps of Photon Index vs.\ $N_{\mathrm{H}}$ for the 2014 (\textit{top}) and 2023 (\textit{bottom}) \rev{datasets fit with a absorbed PL model that accounts for both the Galactic foreground absorption and an absorber at redshift $z=4.715.$}. The contours for the 2014 data are indicated by a dashed line, while the 2023 contours are shown with a solid line. From inside to outside, the lines mark the $1\sigma$, $2\sigma$, and $3\sigma$ contours. \rev{We include results from previous work and mark the value of the Galactic absorption column with a dash-dotted line.}}
    \label{fig:conf_maps}
\end{figure}

We start by assessing the absorption in the spectra taken with contemporaneous observations in 2014 and 2023. Due to the \swift-XRT exposures being very short, we include all data taken within several days of the \nustar observation under the assumption that no rapid change occurs for the spectral shape or absorption.
For the 2014 epoch, we include two \swift-XRT pointings with a combined exposure of 9.6\,ks, and for the 2023 epoch, we choose seven \swift-XRT observations yielding a combined exposure of 18.4\,ks in the soft X-ray band.
We model the spectra with a \texttt{constant*tbabs*}\textbf{ztbabs*}\texttt{cflux(powerlaw)} model, in which the constant is used to counterbalance slight calibration offsets between the \nustar detectors FPMA and FPMB and \swift-XRT. \rev{\texttt{ztbabs} is used to account for source-intrinsic cold absorption at $z=4.715$.}
We consider data taken between 0.3 and 10\,keV for the XRT spectra and from 3 to 80\,keV for the \nustar spectra.
The resulting best fit parameters are listed in Table~\ref{tab:xray_fitresults_nh_constraint}.
When left as a free parameter, the neutral density absorption $N_\mathrm{H}$ \rev{at redshift 4.715} is found to be significantly higher than the Galactic foreground value. To investigate the correlation between $N_{\mathrm{H}}$ and the photon index, we compute confidence contours for $1\sigma$, $2\sigma$, and $3\sigma$ confidence for two parameters of interest \rev{that we show in} Fig.~\ref{fig:conf_maps}. \rev{In the plot, we include results from previous analyses by \citet{Boller2000}, \citet{Fabian2001}, and \citet{Worsley2004} based on data collected in the soft X-ray range by \textit{ROSAT}, \textit{Beppo-SAX}, and \xmm\ between 1998 and 2002. On the 2014 data, our spectral fit provides a less constraining absorbing column measurement than the analysis in \citet{Paliya2016}. A scenario with no absorption beyond the Galactic column density is excluded only at 95\% confidence.}.

Comparing the best-fit results from both epochs, the X-ray flux (as measured with an absorbed power law) is about 50\% higher in 2023 than in 2014, which is likely linked to the \gm-ray flare.
The difference between both fluxes is on the order of $>14\sigma$.
In addition, the spectrum taken in 2023 is significantly harder \rev{while showing the same amount of} absorption \rev{as} the spectrum from 2014.

\begin{figure*}
    \centering
    \includegraphics[width=0.49\textwidth]{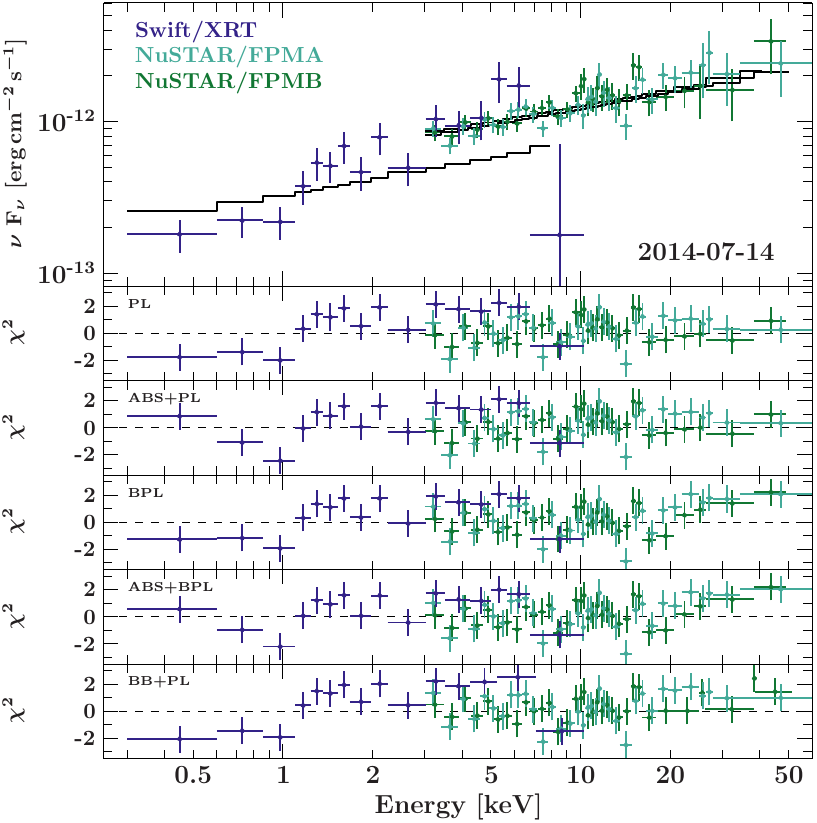}\hfill
    \includegraphics[width=0.49\textwidth]{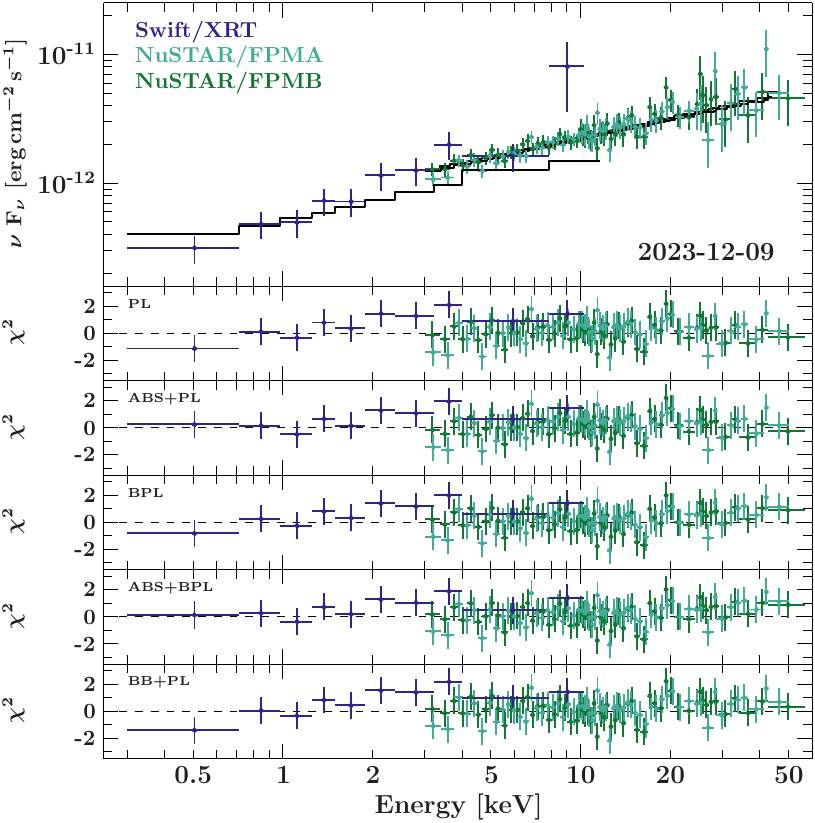}
    \caption{Broadband X-ray spectra of the observation taken in 2014 (\textit{left}) and 2023 (\textit{right}). In each top panel we show the spectral bins and the power law model with foreground absorption. The residual panels are shown for each model we apply and list in Table~\ref{tab:xray_fit_model_comparison}.}
    \label{fig:xray_spectra}
\end{figure*}

We further test \rev{whether other} model\rev{s can} describe the broadband X-ray spectra \rev{equally well or better}. Due to some variability present in the soft X-ray range, we now only fit strictly simultaneous data together. 
We test f\rev{ive} different models: an absorbed power law with the absorption column fixed to the Galactic foreground value (PL), an absorbed power law with \rev{an additional} absorption parameter \rev{at redshift} $z=4.715$ free to vary (ABS+PL), a broken power law with the absorption fixed on the Galactic value (BPL), a broken power law with the $z=4.715$ absorption parameter being free (ABS+BPL)\rev{, and a blackbody plus power law model (BB+PL)}. \rev{The PL model serves as our baseline, since blazars in the local Universe typically do not show a deviation from a power law with Galactic foreground absorption. We expect the results from the ABS+PL model to match the dedicated fit described in the previous paragraph, however, with slightly looser constraints given that we use only strictly simultaneous data here.
The BPL model represents emission from an electron distribution with a spectral cooling break or a scenario in which two distinct processes (e.g., SSC and EC) contribute to the X-ray spectrum, with the break $E_b$ indicating the energy where the transition occurs. ABS+BPL considers an additional intrinsic absorption on top of the broken power law. Lastly, the motivation for testing a BB+PL model originates from the predictions of bulk Comptonization of thermal photons from the BLR \citep{celotti2007}}.

The best-fit results for each model are listed in Table~\ref{tab:xray_fit_model_comparison} \rev{and the spectral plots including residuals for each model are shown in Fig.~\ref{fig:xray_spectra}}. Based \rev{solely} on the $F$-statistics and $\textit{p}$-values, we find that for both epochs the \rev{BPL model} yields the best fit, although \rev{for} the 2014 data, the \rev{ABS+BPL model} gives equally good results.
\rev{The BB+PL model is an improvement compared to the PL model as well, however the fraction of the total contribution to the model by the black body compared to the power law is only on the order of 2\% for both data sets.
Interestingly, the ABS+PL model is an improvement compared to the PL model for the 2014 data set (\textit{p}-value} $<0.05$ \rev{), but not for the 2023 data set. However, we note that the statistics for the 2023 data already yield a satisfactory fit to the spectrum with a reduced $\chi^2<1$.}
\rev{Discriminating features between the tested spectral models appear most prominently in the soft X-ray band, covered only by \swift. Due to higher photon statistics and better signal to noise, the cost function (C-statistic or $\chi^2$) in standard minimization algorithms is dominated by the \nustar data, favoring models that fit well at higher energies even if they represent a suboptimal fit of the soft X-ray spectrum. Therefore, model selection based solely on global fit statistics can be misleading.}

\rev{In the following, we examine the residuals under different models (see Fig.~\ref{fig:xray_spectra}) to evaluate their performance in describing the soft X-ray spectrum of \bzq. The 2014 data set generally shows larger residuals compared to the 2023 data set.
For the latter, we find that the ABS+PL, BPL, and ABS+BPL models sufficiently describe the slope of the \swift-XRT data, with those models including the strong intrinsic absorption yielding better agreeing residuals at the lowest energies. For the BB+PL model, we do not see an improvement in the residuals compared to the PL model in either data set.
We conclude that we cannot significantly distinguish between ABS+PL, BPL, and ABS+BPL.}

When applying the broken power law model, the flux in the full 0.3--80\,keV band decreases 
by 50\% and 25\% in the 2014 and 2023 epoch, respectively, but is still consistent with a significant increase by a factor of $\sim2.5$ between the two epochs.
While for both data sets the energy of the spectral break is consistently at $\sim20$\,keV, the photon index $\Gamma_2$ for the higher energies cannot be constrained for the 2014 data set (we have limited the maximum value of $\Gamma$ to 4).

\citet{Paliya2016} initially presented the simultaneous \swift-XRT and \nustar from 2014 and reported that both a power law with additional absorption on the order of $\sim7\times10^{22}\,\mathrm{cm}^{-2}$ or a broken power law with a break at $\sim$5\,keV describe the data equally well, and significantly better than a power law with only Galactic foreground absorption.
Our results \rev{agree with} their findings \rev{with regard to the absorption strength and flux, but we obtain a softer photon index and for the broken power law} model \rev{we} find the break in the spectrum at much higher energies.

\begin{table*}
\centering
\caption{Model comparison for the 2014 and 2023 epoch using only simultaneous data. \rev{For the BPL and the ABS+BPL models, we list the break energy} $E_{\mathrm{b}}$\rev{, and for the BB+PL model we list the peak temperature} $kT$.}\label{tab:xray_fit_model_comparison}
\centering
\resizebox{1.05\textwidth}{!}{
\begin{tabular}{lcccccccc}
\hline
\multicolumn{9}{c}{2014} \\ 
Model   & $N_{\mathrm{H}}$  &  $\Gamma_1$ &  $\Gamma_2$  &  $E_{\mathrm{b}}$ / $kT$  & Flux$_{0.3-80\,\mathrm{keV}}$   & $\chi^2$ / d.o.f. & F-stat & \textit{p}-value   \\ 
   & [$\mathrm{cm}^{-2}$]  &  &    &  [keV]  & [$10^{-12}$\,erg\,$\mathrm{cm}^{-2}$\,$\mathrm{s}^{-1}$]   &  & &   \\ \hline
PL      &  $9.29\times10^{19}$   & $1.65\pm0.04$  &  --  &  --   & $5.15\pm0.21$  & 429.5 / 346 & -- &  -- \\
ABS+PL  & $6.0^{+4.2}_{-2.8}\times10^{\rev{22}}$   & $1.6\rev{7}\pm0.04$  & --   &   -- & $5.0\rev{5}\pm0.2\rev{1}$ & \rev{422.42 / 345} & 5.8\rev{0}  & 0.016  \\
BPL     & $9.29\times10^{19}$  & $1.55^{+0.05}_{-0.04}$ & $>3.2$  & $19.4^{+2.4}_{-2.3}$ & $3.24\pm0.18$  & 392.7 / 344 & 16.12 & $2.0\times10^{-7}$ \\
ABS+BPL & $4.2^{+3.4}_{-2.5}\times10^{\rev{22}}$ & $1.5\rev{9}\pm0.05$ & $>3.\rev{3}$ & $20.\rev{5}^{+1.9}_{-2.\rev{7}}$ & $3.31\pm0.1\rev{7}$  & \rev{389.04 / 343} & 11.\rev{90} & $\rev{2.0}\times10^{-7}$ \\ 
\rev{BB+PL}     & $9.29\times10^{19}$  & $\rev{1.74}^{+0.08}_{-0.07}$ & --  & $\rev{3.1}\pm\rev{0.5}$ & $\rev{4.30}^{+0.29}_{-0.27}$  & \rev{412.72 / 344} & \rev{6.99} & $\rev{1.05}\times10^{-3}$ \\\hline
\multicolumn{9}{c}{2023} \\
Model   & $N_{\mathrm{H}}$  &  $\Gamma_1$ &  $\Gamma_2$  &  $E_{\mathrm{b}}$ / $kT$  & Flux$_{0.3-80\,\mathrm{keV}}$   & $\chi^2$ / d.o.f. & F-stat & \textit{p}-value  \\ 
   & [$\mathrm{cm}^{-2}$]  &  &    &  [keV]  & [$10^{-12}$\,erg\,$\mathrm{cm}^{-2}$\,$\mathrm{s}^{-1}$]   &  & &  \\ \hline
PL      & $9.29\times10^{19}$  & $1.500^{+0.026}_{-0.025}$ &  --  &  -- & $10.3\pm0.3$ & 275.7 / 369 & -- & -- \\
ABS+PL  &  $\rev{3.7}^{+3.8}_{-2.8}\times10^{\rev{22}}$  & $1.506\pm0.026$  &  -- & -- & $10.3\pm0.3$ & \rev{273.80 / 368} & 2.\rev{55}  & 0.1\rev{1}  \\
BPL     & $9.29\times10^{19}$  & $1.44\pm0.04$ & $2.21\pm0.26$ & $20.3^{+1.3}_{-1.8}$ & $8.3\pm0.6$  & 261.4 / 367 & 10.04  & $5.7\times10^{-5}$\\
ABS+BPL & $\rev{2.5}^{+3.6}_{-2.5}\times10^{\rev{22}}$ & $\rev{1.45}\pm0.04$ & $\rev{2.20}\pm0.26$ & $20.\rev{4}^{+1.\rev{3}}_{-1.\rev{9}}$  & $8.3\pm0.6$  & \rev{260.59 / 366} & \rev{7.07} & $1.\rev{3}\times10^{-4}$\\
\rev{BB+PL}     & $9.29\times10^{19}$  & $\rev{1.57}\pm\rev{0.06}$ & --  & $\rev{4.1}^{+0.8}_{-0.9}$ & $\rev{9.2}\pm\rev{0.5}$  & \rev{266.48 / 367} & \rev{6.35} & $1.95\times10^{-3}$ \\\hline
\end{tabular}
}
\end{table*}

\subsection{X-ray variability}\label{sec:xrayvar}
\begin{figure*}
    \includegraphics[width=0.944\textwidth]{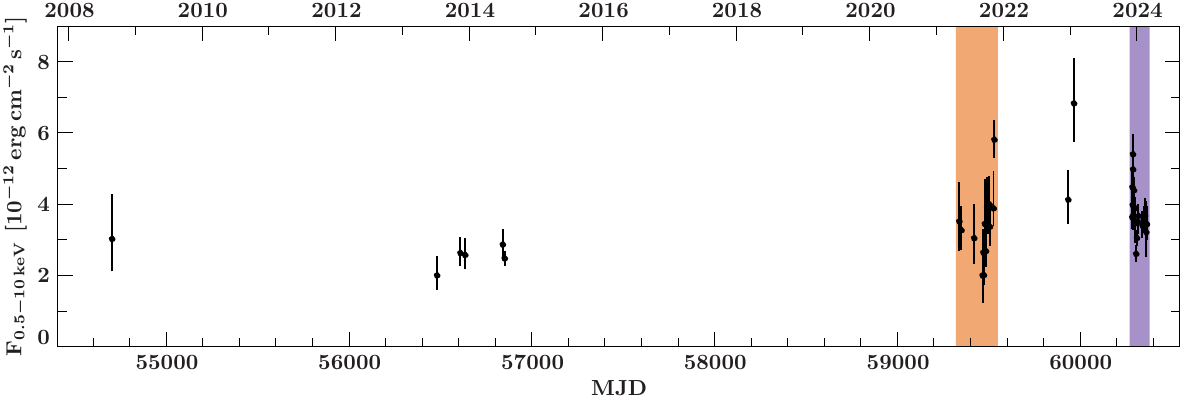}
    \includegraphics[width=0.99\textwidth]{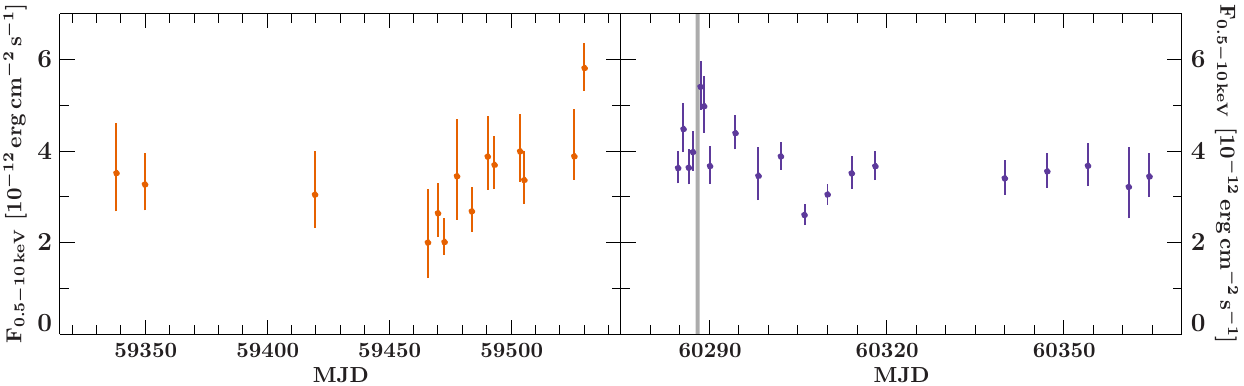}
    \caption{\swift-XRT light curves. Top: full LC from 2008 to 2024. Bottom: Zoom in into the denser monitoring in 2021 and 2024, where the later time range covers the time range during which the gamma-ray flare occured. The time of the \nustar DDT observation is marked by a grey shaded region in the bottom right light curve.}
    \label{fig:xray_lcs}
\end{figure*}

Based on the 2014 and 2023 data set collected for the full X-ray band, we can confirm that the photon index as well as the flux have changed between the two epochs. 
In order to produce a light curve, we model each spectrum with a simple power law including Galactic foreground absorption (i.e., fixing $N_\mathrm{H}$), however, we exclude all data for which the exposure time is less than 0.5\,ks or if the source spectrum contains less than 50\,counts. In addition, observations taken within 24\,hours of each other are fit simultaneously. We use C-statistics \citep{cash1979} due to the low number of photons in the majority of the \swift-XRT spectra.
The best-fit results are listed in Appendix Table~\ref{tab:xrt_lc_fitresults} and the light curve is shown in Fig.~\ref{fig:xray_lcs}.

Using the monitoring time ranges from 2021 May through 2021 November as well as December 2023 through February 2024, we can also study shorter term variability.
In particular, as seen in Figure~\ref{fig:xray_lcs}, our monitoring in 2023/2024 contains daily snapshots for the initial seven days after the \gm-ray flare, followed by seven observations at four-day cadence. After a gap of 22 days, we monitored \bzq with weekly cadence.
We test for flux variability by fitting each light curve section to a constant flux, deriving $p$-values for the constant flux hypothesis using a $\chi^2$ test. 
We find $p=9.4\times10^{-6}$ and $p=4.5\times10^{-5}$ for the 2021 and the 2023/2024 data sets, respectively, which confirms that variability is present in \bzq in the 0.3--10\,keV band over timescales of several days \rev{in our observer's frame, and a factor of $1/(z+1)$ shorter in the restframe of \bzq}.
To further quantify the variability time scales, we perform a standard calculation of exponential flux doubling/halving times between each flux point in each epoch, respectively. Due to the rather large uncertainties on the flux, we can only obtain one significant halving time for a flux point combination in 2023, for which we find $\tau_{\mathrm{h}}=\rev{1.2}\pm\rev{0.4}$ days ($\sim3.4\sigma$) \rev{in the source rest frame}. 
Additionally, we find \rev{the most rapid and $>2\sigma$} doubling time of \rev{$12\pm6$\,hours} in the 2023/2024 epoch \rev{as well}.

\subsection{Optical polarization}
\label{sec:polarization}
The optical and infrared radiation from gamma-ray emitting blazars at $z\gtrsim 3$ is generally attributed to thermal emission from the AGN accretion disk \citep[e.g.,][]{Paliya2016,Paliya2019,Liao2019,Marcotulli2020,Gokus2024}. This interpretation is supported by two main factors: first, the high redshift shifts the disk emission, typically in the blue and ultraviolet bands in low-$z$ AGN, towards longer wavelengths. Second, the larger estimated or inferred black hole masses in high-$z$ blazars would result in lower temperature disks \citep[$T_{\text{disk}}\propto M_{\text{BH}}^{-1/4}$,][]{Shakura1973}, further shifting the disk emission towards the optical and infrared bands. 

The moderate R band polarization $\mathrm{P}=(8.6\pm3.0)$\% measured from \bzq indicates that this source may not conform to this scenario, at least during the gamma-ray flare examined in this study. Moderate to high levels of polarization ($3\% \lesssim P \lesssim 30\%$) are typically expected from synchrotron radiation, with higher polarization levels indicating a more ordered magnetic field in the emitting region. In contrast, thermal emission from the accretion disk is expected to be unpolarized. Combined spectral and polarization surveys of blazars converge in considering objects with $P\gtrsim 3\%$ as synchrotron-dominated \citep{Smith2007}. 

To further assess whether the observed R-band polarization from \bzq can distinguish between a synchrotron and thermal origin of the optical emission, we examine publicly-available optical polarization observations for other gamma-ray bright blazars \citep[][\footnote{\url{https://james.as.arizona.edu/~psmith/Fermi/}}]{Smith2016}. We observe that objects with SEDs where optical emission is dominated by thermal disk emission or galactic foreground (Mrk~501, 3C~273, 1ES~2344+514) show optical polarizations $\leq 3\%$. Objects for which the optical SED is dominated by continuum emission from synchrotron radiation (OJ~287, 3C~279) show $P\gtrsim 10\%$. Finally, objects for which the optical emission is dominated by synchrotron continuum but show some contribution from a thermal disk component (PKS~1222+216, PKS~1510-089) show variable polarization in the range $3\% \lesssim P \lesssim 6\%$. The observed $\mathrm{P}=(8.6\pm3.0)$\% in \bzq suggests that the R-band emission of the quasar being dominated by synchrotron radiation from the jet, with the possibility of some contribution from thermal emission from the AGN disk.

\subsection{Broadband SED Modeling}
\label{sec:modeling}
\begin{figure*}
    \centering
    \includegraphics[width=\textwidth]{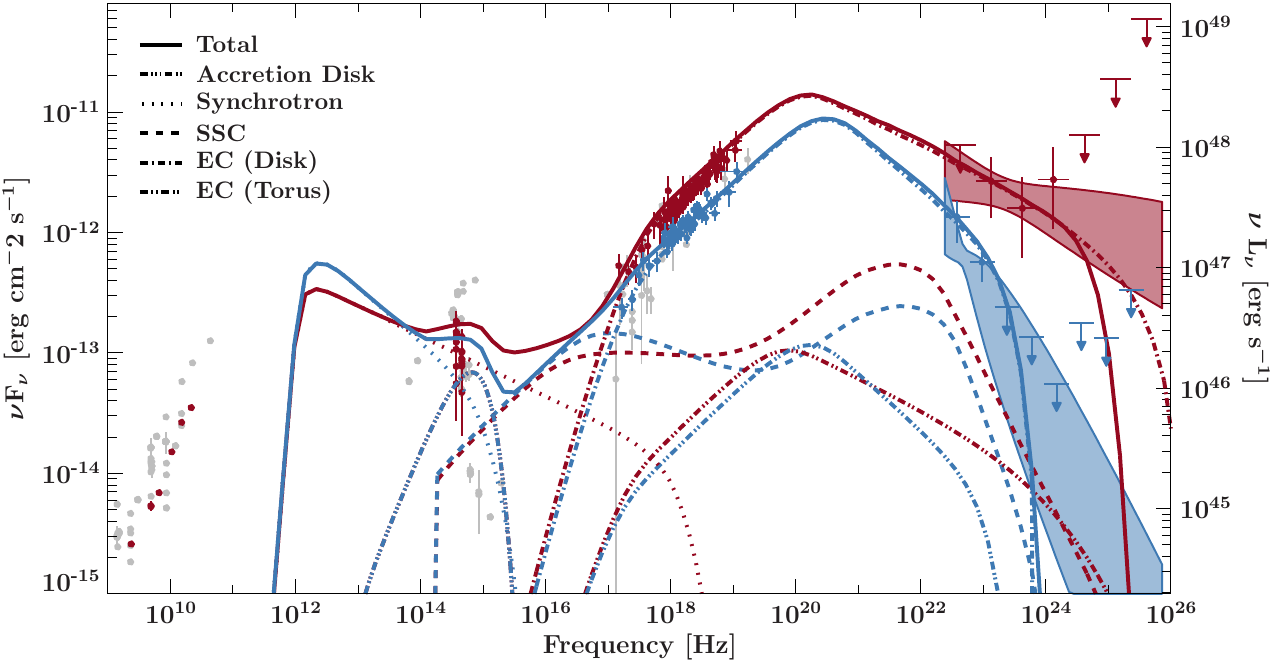}
    \caption{Broadband SED of \bzq for non-simultaneous, archival multiwavelength data (blue), and for the active state in 2023 with contemporaneous data (red), modeled with the stationary one-zone leptonic model described in \cite{Boettcher13}.}
    \label{fig:broadband_sed}
\end{figure*}

We compiled quasi-simultaneous SEDs for both the low state in 2014 and the flaring state in 2023, which are shown in Fig.~\ref{fig:broadband_sed} with blue and red symbols, respectively. These data were complemented by archival radio, IR, optical, and X-ray data (black symbols). Due to the sparse simultaneous IR--optical coverage, the low-frequency component of the SED is very poorly constrained. However, the  moderate polarization  of $8.6 \pm 0.3$\% during the 2024 high state suggests that jet synchrotron radiation makes a non-negligible contribution to the optical flux during that period (see \S \ref{sec:polarization}). This assumption guided our choice of parameters to fit the SEDs, using the stationary, single-zone leptonic model developed by  \cite{Boettcher13}. 

The \cite{Boettcher13} leptonic model assumes a spherical emission region of radius $R_b$ propagating along the jet with bulk Lorentz factor $\Gamma$. It is magnetized with a magnetic field of strength $B$. A non-thermal, relativistic electron population is injected by an unspecified rapid  acceleration mechanism, with spectral index $q$ and low- and high-energy cut-offs at $\gamma_\mathrm{min}$ and $\gamma_\mathrm{max}$, respectively. The code then evaluates an equilibrium distribution resulting from the balance between injection, radiative  cooling, and escape from the emission region on an energy-independent escape time scale $t_\mathrm{esc} = \eta_\mathrm{esc} \, R_b/c$. Radiation mechanisms included in the code are synchrotron radiation, synchrotron-self-Compton (SSC), and external-Compton (EC) from  scattering of the direct accretion-disk radiation, EC (disk) and an isotropic (in the AGN rest frame) thermal radiation field, in this case with parameters appropriate for IR emission from the dust torus, EC (DT). 
We use this model to perform a fit-by-eye and obtained acceptable fits using the parameters listed in Table~\ref{tab:SEDfitparameters}. Guided by the hardening of the \fermi-LAT spectrum from  2014 to 2023,  the most significant parameter changes between the two states is a significantly harder electron injection spectrum, extending to much higher energies during the flaring state. Note, however, that the SED, especially the low-frequency component, is poorly constrained, resulting in significant parameter degeneracies. 
Therefore, the values listed in Table~\ref{tab:SEDfitparameters} should be considered as constituting a plausible, but not unique, scenario that can describe the SED in the framework of a single-zone leptonic emission model.
For example, we note that the EC (DT) component is subdominant, with the high-energy radiation being largely described by EC (disk) and SSC emission. Other radiation fields, such as the BLR, could also contribute to the observed \gm-ray radiative output.
As is commonly found in blazars, the high-energy emission region modeled here is optically thick to synchrotron self-absorption in the radio band. Therefore, the radio emission is expected to be produced in the larger-scale jet, not modeled here. 

\begin{table}
\centering
\caption{SED model fit parameters used for the fits shown in Fig. \ref{fig:broadband_sed}. $q=$ electron spectral index at injection. $d=$ injection height. }\label{tab:SEDfitparameters}
\centering
\begin{tabular}{lcc}
\hline
Parameter & Long-term & Flare \cr 
\hline 
$L_e$ [erg s$^{-1}$] & $4.7 \times 10^{45}$ & $5.7 \times 10^{45}$ \cr 
$\gamma_\mathrm{min}$ & 100 & 60 \cr
$\gamma_\mathrm{max}$ & $2.5 \times 10^3$ & $1 \times 10^5$ \cr 
$q$ & 2.9 & 2.5 \cr 
$B$ [G] & 8.0 & 7.0 \cr 
$\eta_\mathrm{esc}$ & 10 & 10 \cr 
$d$ [pc] & 0.12 & 0.1 \cr 
$\Gamma$ & 14.6 & 14.6 \cr 
$L_\mathrm{disk}$ [erg s$^{-1}$] & $2 \times 10^{47}$ & $2 \times 10^{47}$ \cr 
$R_b$ [cm] & $8 \times 10^{15}$ & $8 \times 10^{15}$ \cr 
$M_\mathrm{BH}$ & $1.7 \times 10^9$ & $1.7 \times 10^9$ \cr 
$T_\mathrm{DT}$ [K] & 1000 & 1000 \cr 
$u_\mathrm{DT}$ [erg\,cm$^{-3}$] & $6.0 \times 10^{-4}$ & $6.0 \times 10^{-4}$ \cr 
\hline 
$L_B$ [erg\,s$^{-1}$] & $3.3 \times 10^{45}$ & $2.5 \times 10^{45}$ \cr 
$\epsilon_{Be}$ & 0.691 & 0.435 \cr 
\hline 
\end{tabular}    
\end{table}

\section{Discussion}\label{sec:discussion}

\subsection{Broadband emission} 
We made use of existing measurements of some properties of \bzq as starting points to derive our SED model parameters.
The reported mass estimates for this source range from ${\sim}1.7\times10^9\,M_{\odot}$ to ${\sim}7.4\times10^9\,M_{\odot}$ \citep[e.g.,][]{shen2011, diana2022}, where the former is derived from the dispersion of the \ion{C}{4} line, and the latter is obtained through measuring the full width half maximum of said emission line.
We find that the expected accretion disk emission from a $1.7\times10^9$\,M$_{\odot}$ black hole, in agreement with values reported in the literature, can describe the observed optical to infrared SED. 
Our choice of the bulk Lorentz factor, $\Gamma=14.6$, was motivated by a 22-year long radio campaign using very-large baseline interferometry \citep{zhang2020}.
Our polarization measurement in the optical R band helps us to determine how much of the emission in this band can be attributed to synchrotron emission, since the measured polarization of $8.6$\% is incompatible with a pure accretion-disk spectrum. As a result, our assumption of the amount of synchrotron emission also led to our choice for the magnetic field strength and to achieve equipartition between the B-field and the relativistic leptons.

Some studies have modeled the broadband SED of \bzq in the past. 
\citet{ghisellini2009} used a leptonic model with a variable distance of the dissipation region for the different states of the source based on X-ray data collected with \textit{Beppo-SAX}, \textit{ASCA}, and \xmm. They find that the dissipation region lies within the BLR at about 1500 Schwarzschild radii, $R_{\mathrm{S}}$, from the central SMBH. For both states that we model with more recent X-ray data and including \gm-ray data from LAT, we find that the dissipation region lies even closer to the SMBH, at $\sim740\,R_{\mathrm{S}}$ during the long-term average state, and at ${\sim}610\,R_{\mathrm{S}}$ during the flaring event in 2023.
Using the same model as \citet{ghisellini2009}, but using a newer X-ray data set from 2014, \citet{Paliya2016} modeled broadband data similar to our long-term SED. 
While we use similar values for the black hole mass, the bulk Lorentz factor, and the accretion disk luminosity, 
the major difference between both models is that \citet{Paliya2016} describe a magnetically dominated jet while we model \bzq under the assumption of equipartion.
Comparing the properties of the class of high-$z$ redshift blazars to our results from the SED modeling, we find that some parameters derived for \bzq seem to lie at the higher end of the range reported by \citep{Paliya2020}, in particular the accretion disk luminosity, the bulk Lorentz factor, and the magnetic field strength.
The long-term average \gm-ray luminosity and photon index of \bzq are in very good agreement with the values found for other \gm-ray detected blazars at $z>3$ \citep{ackermann2017}.

\subsection{Comparison with other luminous blazars}

\begin{figure}
    \centering
    \includegraphics[width=0.45\textwidth]{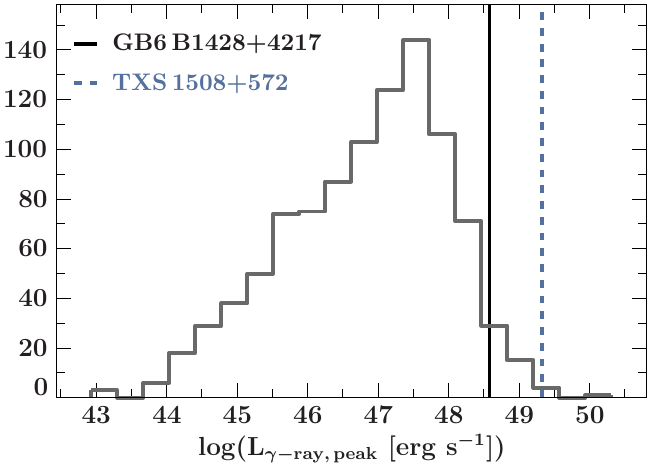}
    \caption{Distribution of peak \gm-ray luminosities obtained from the 4FGL-DR4 with redshift information taken from 4LAC-DR3. The histogram is computed for a number of 20 bins between the minimum and maximum luminosities with equal spacing in the logarithmic space. We mark the peak \gm-ray luminosities for \bzq and TXS\,1508+572 \citep[see][]{Gokus2024} with solid and dashed lines, respectively.}
    \label{fig:hist_gamma_maxlum}
\end{figure}

To put the luminosity of the \gm-ray flare from \bzq in context, we use the peak yearly fluxes and time-averaged \gm-ray photon indices from the {\it Fermi} 4FGL-DR4 catalog \citep{4fgl-dr4} with redshift estimates from 4LAC-DR3 \citep{4lac-dr3} to calculate k-corrected peak luminosities for a sample of of 977 \gm-ray detected blazars. These peak luminosities may be underestimated compared to shorter timescale \gm-ray flares reported in the literature. The resulting peak luminosity distribution is shown in Fig.~\ref{fig:hist_gamma_maxlum}. 
The flare luminosity of \bzq of $(3.9 \pm 3.8) \times 10^{48}$\,erg\,s$^{-1}$ puts it in the top 5\% of most luminous flares seen by \fermi. 

The dimensionless Compton Dominance ($CD$) parameter, defined as the ratio between the peak synchrotron luminosity, $L_\mathrm{syn}$, and the peak IC luminosity, $L_\mathrm{IC}$,
can be used as a proxy for the relative energy densities of external radiation and  magnetic fields \citep[e.g.,][]{pacciani2014}:
\begin{equation}
    {CD}= \frac{L_C}{L_\mathrm{syn}} = \frac{U'_\mathrm{ext}}{U'_B} = \frac{U_\mathrm{ext}\Gamma^2_\mathrm{bulk}}{U'_B}
\end{equation}
Based on the peak values of the low- and high-energy component listed in the 4LAC-DR3 \citep{4lac-dr3}, we compute $CD$ values based on the average SED for a large sample of blazars. 
\bzq displays a high Compton dominance of $CD\sim 15$ in an average SED state, increasing to $CD\sim 40$ during the 2023 \gm-ray flare.

We find five sources in the same peak luminosity bin as \bzq that have been studied in detail in the literature: PKS~0346$-$27, PKS~0402$-$362, PKS~0528+134, TXS~0536+145 and PKS~2023$-$07, with redshifts in the range $0.99 \leq z \leq 2.69$. 
\bzq is most similar to PKS\,0346$-$27, which shows a high $CD\sim 25$ in its average SED state \citep{angioni2019,kamaram2023}, increasing to $\sim 45$ during flares \citep{angioni2019}\footnote{We note that the Compton dominance is not explicitly listed in other publications. Hence, we calculate these values from reading the peak fluxes of the SED figures.}. The other four objects show a smaller Compton dominance, with their time-averaged SEDs yielding $CD \lesssim 5$.

\subsection{High-redshift blazars within the blazar population}
\begin{figure}
    \centering
    \includegraphics[width=0.49\textwidth]{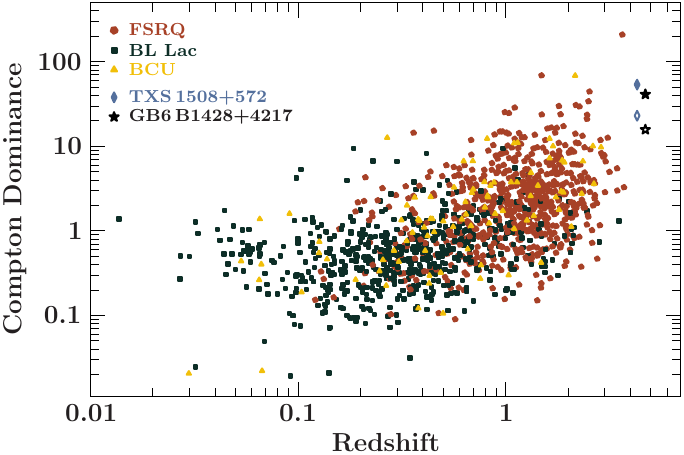}
    \caption{Compton dominance over redshift for blazars with a known redshift and Compton peak listed in the 4LAC-DR3 \citep{4lac-dr3}. We include the Compton dominance of TXS\,1508+572 (diamond symbol) and \bzq (star symbol) for both the quiet (empty symbol) and active (filled symbols) states to illustrate the significant change during flaring episodes.}
    \label{fig:redshift_vs_CD}
\end{figure}

For a comparison with the entire blazar class, we use the 4LAC-DR3 and plot the average Compton dominance as a function of redshift of the 4LAC blazars for which a redshift is known in Fig.~\ref{fig:redshift_vs_CD}.
At higher redshifts, we are biased toward detections of more luminous sources due to the limited sensitivity of \fermi-LAT, hence we are likely missing detections of blazars with a small Compton dominance in the early Universe.
53 blazars and blazar-like AGN exhibit a strong Compton dominance (${CD}\geq10$), with a median redshift $z=1.55$ for these sources. Only ten sources at $z<1$ (i.e., within the last $\sim6$\,billion years) belong to this group, while 15 fall into the time of cosmic noon ($z>2$).

We add \bzq and TXS\,1508+572 \citep{Gokus2024} to Fig.~\ref{fig:redshift_vs_CD} with distinct symbols 
showing their Compton dominance during quiet-state/average SED (unfilled symbols) and flaring states (filled symbols). Both objects show an increase in $CD$ during flares.
The increase of the Compton dominance by a significant factor has also been observed for luminous flares in FSRQ with $z<1$, e.g., 3C\,279 \citep{hayashida2015}, or 3C\,454.3 \citep{zhou2021}, as well as for much less luminous, but blazar-like sources such as \gm-loud Narrow Line Seyfert 1 galaxies \citep[e.g., PKS\,2004$-$447;][]{gokus2021}.
However, on average, most blazars exhibit a low Compton dominance:  based on values derived from the 4LAC-DR3, the average $CD$ for BL Lac objects is 0.9, while it is 3.9 for FSRQs.

\begin{figure}
    \centering
    \includegraphics[width=0.47\textwidth]{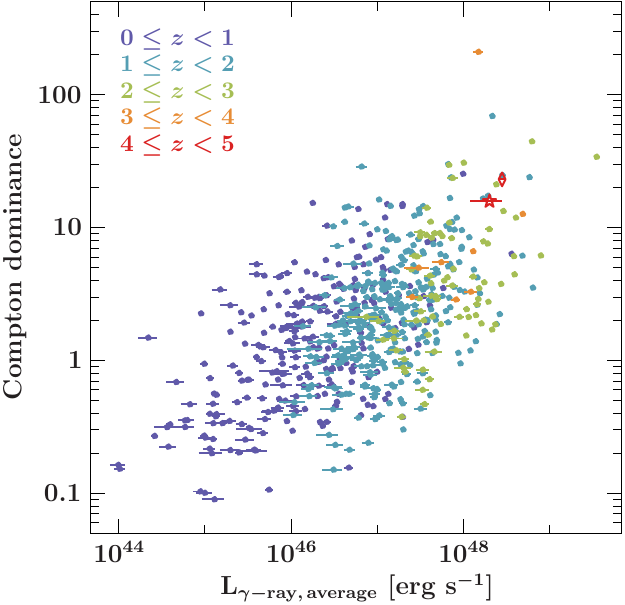}
    \caption{Compton dominance vs.\ average \gm-ray luminosity for all FSRQs with a known redshift and Compton peak listed in the 4LAC-DR3 \citep{4lac-dr3}. TXS\,1508+572 (diamond symbol) and \bzq (star symbol), for which we also show the average \gm-ray luminosities and Compton dominance values, are the only two blazars with $z>4$.}
    \label{fig:lum_vs_CD}
\end{figure}

\citet{2017A&A...606A..44N} discuss in detail that the denser external radiation fields that are responsible for the external IC emission can result from a more luminous accretion disk, indicating that Compton dominance can be used as a proxy for more efficient accretion processes of the central SMBH. Furthermore, they find that Compton dominance scales with the equipartition parameter $\epsilon_{Be}=P_B/P_e$, which for values $<1$ 
indicates that the jet is likely matter-dominated. For \bzq, $0.4<\epsilon_{Be}<0.7$, within that range.
In addition, a correlation between the Eddington ratio and the Compton dominance has been reported by \citet{paliya2021}.

The detection of FSRQ at high redshifts allows us to verify the predictions of the blazar sequence \citep{fossati1998,meyer2011} at the highest luminosities. 
The blazar sequence describes how more luminous blazars tend to have lower synchrotron peak frequencies and higher Compton dominance, driven by stronger radiative cooling via external photon fields. 
Originally built on a radio and X-ray selected sample, the \gm-ray selected \fermi blazar sequence \citep{ghisellini2017} revealed that the shift of the peak frequencies is only seen for BL Lac-type objects, while the increasing \gm-ray luminosity (and with it an increased Compton dominance) is solely present for FSRQs. Other work even questions the validity of the blazar sequence altogether, and it has been argued that the correlation of the peak frequency and peak luminosity is an observational selection effect \citep[e.g.,][]{nieppola2008, giommi2012, fan2017, keenan2021}. A detailed review on this topic is given by \citet{prandini2022}.
In the framework \rev{of the sequence}, very luminous FSRQs have high accretion rates, producing dense external radiation fields (e.g., from the broad line region) that enhance Compton scattering and explain their observed high Compton dominance.
Figure~\ref{fig:lum_vs_CD} shows this correlation for a large sample of detected FSRQ, with values calculated from extracted average fluxes and redshifts from the 4FGL-DR4 and 4LAC-DR3 catalogs. 

Due to the increasing sampling volume, the median luminosity of FSRQs is observed to increase with increasing redshift. At $z>4$, TXS\,1508+572 and \bzq show Compton dominance values of 23 and 16, respectively. While these are among the highest values in the FSRQ sample, they are not outliers but follow the general trend of rising Compton dominance with increasing luminosity observed also at lower redshifts.  
Based on the observed trend, we can infer that gamma-ray detected blazars at high redshift are compatible with being objects with high accretion rates, as expected from the blazar sequence. With higher accretion rates, there is greater available accretion power, resulting in more powerful jets. Additionally, the accretion disk becomes brighter and more radiatively efficient, capable of ionizing surrounding gas such as the broad line region. These dense external photon fields serve as seed photons for Compton scattering, enhancing the radiative cooling rate of the non-thermal electrons in the jet. This process leads to a higher observed Compton dominance. 

Given that the energy density of the cosmic microwave background (CMB) scales as $(1+z)^4$, one might expect Compton scattering on the CMB to significantly contribute to the \gm-ray SED for high-redshift objects. However, the modeling of SEDs and the absence of an excess in Compton dominance for high-redshift objects do not suggest that Compton scattering on the CMB is a significant cooling mechanism. This implies that the dissipation region of the blazar must be located within $<10^6$ gravitational radii from the central black hole, where the infrared dust torus and the broad line region dominate the radiation energy density in the jet environment \citep{ghisellini2009}.

In optical surveys, the number density of quasars is highest at $z \sim 2\text{-}3$ \citep{richards2006}, at the peak epoch of black hole accretion and star formation in the Universe. Our analysis shows that energy dissipation in \gm-ray-emitting FSRQs at $z>4$ seems to follow the same mechanisms observed in luminous FSRQs found in our local universe ($z<1$).

\subsection{Interpretation of anomalous X-ray spectrum}
\rev{We confirm the presence of a peculiar feature in the broadband X-ray spectrum of \bzq in data obtained during a flaring state of the source.}

In earlier data, such as the 1999 \textit{BeppoSAX} spectra reported by \citet{Fabian2001}, a warm absorber with an equivalent column density of $n_\mathrm{H} \sim 3\times10^{22}\,\mathrm{cm}^{-2}$ was suggested to explain the flattening of the X-ray spectrum towards low energies--compared to the higher energy power-law \citep{Boller2000,Fabian2001,Worsley2004,worsley2006}. 
\rev{The expectation under this scenario is that}
the column density of this local warm absorber does not change significantly over time \rev{and} similar column densities \rev{are measured} \rev{throughout time}. \rev{In particular, at a} higher overall X-ray flux, the presence of a warm absorber would likely be seen at higher significance as a result of increased statistics.
Our analysis of the 2023 X-ray spectra \rev{does support this scenario, but we also find that a broken power law (without a need for intrinsic absorption) statistically does describe our X-ray data equally well. Additionally, a high intrinsic absorption would lead to large reddening in the optical band, which is not observed }\citep{Paliya2016}.

\citet{celotti2007} proposed an alternative interpretation of the anomalous soft X-ray spectrum of \bzq, suggesting that the flattening toward low energies could be due to a bulk Compton component. First proposed in \citet{begelman1987}, the bulk Compton process involves cold, thermal electrons in the jet cooling on radiation fields such as disk or BLR emission via Compton scattering, resulting in a soft X-ray excess.
\rev{In this scenario, the spectrum is expected to flatten at soft X-ray energies due to the presence of an additional black body component (bulk Compton) peaking at mid X-ray temperatures \citep{celotti2007}.
As described in Section~\ref{sec:xrayspectra}, we are able to fit the combined \swift and \nustar spectra from 2014 and 2023 with a black body peaking at $kT\sim3-4$\,keV, but the ratio of black body to power law flux is very low, and the residuals do not indicate that the additional black body component significantly improves the spectral fit.}

\rev{The limited photon statistics in the soft X-ray band from \swift\ prevent a high-confidence assessment of features such as local cold absorption or a bulk Compton component. Definitive tests for these scenarios will require high-fidelity soft X-ray spectra (e.g., from \xmm) obtained at different flux states. A bulk Compton feature would be more pronounced at lower levels of external Compton continuum, while absorption by local cold gas would imprint similar spectral signatures regardless of flux level. Simultaneous \nustar\ observations would help constrain the hard X-ray component, improving the sensitivity to resolve anomalous soft X-ray features.}

\section{Summary and Conclusion}
In this work, we have analyzed the high-energy flare from what is currently the most distant \gm-ray blazar \bzq detected by \fermi-LAT.
At X-ray energies, our follow-up campaign consisted of a \nustar observation during the flaring epoch, as well as 19 observations with \swift, with which we monitored the source for over two months.
Optical data in the R and I bands were obtained with telescopes at the Sierra Nevada Observatory and the Perkins telescope. Using the latter, we measured an optical polarization of $\sim8$\%, which confirms that the optical emission during the flare was dominated by synchrotron emission and puts constraints on the luminosity of the accretion disk, which is otherwise poorly constrained for this source.
The flare was accompanied by harder-when-brighter behavior at \gm-ray energies, as is common in FSRQs.
The \gm-ray luminosity of \bzq during the flare places it amongst the top 5\% of luminous flaring events recorded by the \fermi observatory. 
Past X-ray observations of the blazar have revealed peculiar spectral features below 2\,keV, which have been interpreted to be caused by different absorption scenarios as well as bulk Comptonization of thermal photons.
In the combined fit of the \swift-XRT and \nustar spectra obtained during the flaring event, we \rev{also} find evidence for these features\rev{. However, we are not able to conclusively rule out whether these originate from source-intrinsic absorption or a change or transition in the X-ray spectrum.}

The hard X-ray flux of \bzq showed a significant increase during the flare, and the \nustar observation provided essential data to model the high-energy emission, which we attribute to external Compton emission connected to the accretion disk.

This high-$z$ blazar presents itself as a prototypical MeV blazar with its high-energy component peaking at $\sim1$\,MeV in the broadband SED.
The high Compton dominance presented in its SED fits into the correlation of \gm-ray luminosity with Compton dominance of FSRQs. 
Despite the high observed luminosity of $\sim4\times10^{48}$\,erg\,s$^{-1}$ and a Compton dominance of 41, the observed SED, when put in context of the larger FSRQ population, suggests that the particle acceleration and cooling mechanisms at play at $z=4.71$ are the same as observed in local high-power FSRQs.

Flaring events from more than 12 billion years ago can enable us to study early blazars at \gm-ray energies. If blazars show a significant increase in Compton dominance during a flare (such as \bzq or TXS\,1508+572), the chance for detectability with the upcoming Compton Spectrometer and Imager \citep[COSI;][]{tomsick2019, tomsick2023} may increase the currently predicted number of four $z>3$ blazar detections \citep{marcotulli2022}. Based on the average flaring event occurrence among the sample of our $\sim80$ high-$z$ blazars, which is one every 14 months, COSI might be able to detect one or two additional sources.

\begin{acknowledgments}
A.~G. acknowledges funding by NASA through grant 80NSSC24K0813 and the McDonnell Center for the Space Sciences at Washington University in Saint Louis. A.~G. and M.~E. thank Jules Halpern for his evaluation on the different redshift measurements.
The authors thank the \nustar team for granting the DDT observation of \bzq in December 2023.
The IAA team acknowledges financial support through the Severo Ochoa grant CEX2021-001131-S funded by MCIN/AEI/ 10.13039/501100011033 and through grants PID2019-107847RB-C44 and PID2022-139117NB-C44. J.O.-S. and D.M. acknowledge financial support from the project ref. AST22\_00001\_9 with founding from the European Union - NextGenerationEU, the \textit{Ministerio de Ciencia, Innovación y Universidades, Plan de Recuperación, Transformación y Resiliencia}, the \textit{Consejería de Universidad, Investigación e Innovación} from the \textit{Junta de Andalucía} and the \textit{Consejo Superior de Investigaciones Científicas}. J.O.-S. also acknowledges founding from INFN Cap. U.1.01.01.01.009.
The research at Boston University was supported in part by the National Science Foundation grant AST-2108622,  and NASA Fermi Guest Investigator grant 80NSSC23K1507.
This study was based in part on observations conducted using the 1.8m Perkins Telescope Observatory (PTO) in Arizona, which is owned and operated by Boston University.
This research made use of hips2fits,\footnote{https://alasky.cds.unistra.fr/hips-image-services/hips2fits} a service provided by CDS.
In this work, the author made use of a collection of ISIS functions (ISIS scripts) provided by ECAP/Remeis observatory and MIT (\url{https://www.sternwarte.uni-erlangen.de/isis/}).
Partly based on observations with the 100-m telescope of the MPIfR (Max-Planck-Institut für Radioastronomie) at Effelsberg. FE, JH, and MK, acknowledge support from the Deutsche Forschungsgemeinschaft (DFG, grants 447572188, 465409577). JW acknowledges funding through DFG Research Unit (Forschungsgruppe) FOR 5195 -- Relativistic Jets In Active Galaxies.
\end{acknowledgments}

\vspace{5mm}
\facilities{\fermi-LAT, \swift-XRT, \nustar, Perkins, Sierra Nevada, Effelsberg}

\software{ISIS, HEASOFT}



\appendix

\section{Best-fit values from X-ray analysis}
In Table~\ref{tab:xrt_lc_fitresults}, we list all analyzed \swift-XRT observations and best fit results that have been plotted in the light curve shown in Fig.~\ref{fig:xray_lcs}.

\begin{table*}[]
\caption{Best fit results from \swift-XRT fits with \texttt{tbabs*\rev{ztbabs*}pegpwrlw}\rev{, with fixed absorption (}$N_{\mathrm{H, Gal.}}=9.29\times10^{19}$\,cm$^{-2}$ and $N_{\mathrm{H, z}}=6.0\times10^{22}$\,cm$^{-2}$).}\label{tab:xrt_lc_fitresults}
\begin{tabular}{llcccccc}
\hline\hline
MJD   & ObsID & Counts  & Exposure & Flux$_{0.3-10\,\mathrm{keV}}$   & Photon Index   & Cash Statistics& Exp. C value + var  \\
   &  &  & {[}ks{]}    & [$10^{-12}$\,erg\,cm$^{-2}$\,s$^{-1}$]   &    &&   \\ \hline
54702.9 & 00036798001 & 26  & 0.5 & $3.0^{+1.3}_{-1.0}$    & $0.9\pm0.4$            & 25.89  & 17.39$\pm$5.17   \\
56480.8 & 00036798002 & 40  & 1.1 & $2.0^{+0.6}_{-0.5}$    & $1.32^{+0.26}_{-0.27}$ & 28.64  & 38.42$\pm$7.13   \\
56609.1 & 00036798006 & 74  & 1.4 & $2.6^{+0.5}_{-0.4}$    & $1.59\pm0.20$          & 78.64  & 61.42$\pm$9.39   \\
56634.9 & 00036798007 & 61  & 1.2 & $2.6\pm0.5$            & $1.58^{+0.22}_{-0.21}$ & 68.89  & 52.94$\pm$8.63   \\
56840.6 & 00080752001 & 104 & 2.2 & $2.9^{+0.5}_{-0.4}$    & $1.30\pm0.17$          & 69.04  & 97.69$\pm$11.26  \\
56851.0 & 00080752002 & 295 & 7.4 & $2.48^{+0.23}_{-0.21}$ & $1.31\pm0.10$          & 187.85 & 227.60$\pm$17.89 \\\hline
59337.2	&	03110986004	&	8	&	0.3	&	\multirow{3}{*}{$3.5^{+1.2}_{-0.9}$}	&	\multirow{3}{*}{$1.2\pm0.4$}	&	\multirow{3}{*}{53.72}	&	\multirow{3}{*}{34.01$\pm$7.05}	\\
59338.2	&	03110986005	&	20	&	0.3	&		&		&		&		\\
59339.2	&	03110986012	&	18	&	0.1	&		&		&		&		\\ \hline
59349.9 & 03111053001 & 49 & 0.9 & $3.3^{+0.7}_{-0.6}$    & $1.54\pm0.23$          & 40.88 & 46.21$\pm$8.08 \\
59419.6 & 03111053003 & 36 & 0.7 & $3.0^{+1.0}_{-0.8}$    & $1.0\pm0.4$            & 25.15 & 33.15$\pm$6.45 \\
59465.9 & 03111053004 & 16 & 0.5 & $2.0^{+1.2}_{-0.8}$    & $0.8\pm0.5$            & 6.41  & 16.06$\pm$4.69 \\
59470.0 & 03111053006 & 36 & 0.8 & $2.6^{+0.7}_{-0.6}$    & $1.52\pm0.27$          & 29.49 & 34.42$\pm$6.70 \\
59472.7 & 03111053007 & 68 & 1.9 & $2.01^{+0.53}_{-0.28}$ & $1.45^{+0.20}_{-0.26}$ & 41.78 & 61.85$\pm$9.11 \\
59477.8 & 03111053008 & 27 & 0.8 & $3.5^{+1.3}_{-1.0}$    & $0.7\pm0.4$            & 24.56 & 25.57$\pm$5.80 \\
59483.9 & 03111053009 & 63 & 1.4 & $2.7^{+0.6}_{-0.5}$    & $1.40^{+0.22}_{-0.21}$ & 46.98 & 57.88$\pm$8.82 \\
59490.4 & 03111053012 & 53 & 0.9 & $3.9^{+0.9}_{-0.8}$    & $1.17\pm0.23$          & 33.04 & 50.40$\pm$8.12 \\
59493.2 & 03111053013 & 84 & 1.3 & $3.7^{+0.7}_{-0.6}$    & $1.41^{+0.19}_{-0.18}$ & 72.75 & 70.03$\pm$9.72 \\
59503.7 & 03111053014 & 65 & 1.2 & $4.0^{+0.9}_{-0.7}$    & $1.13\pm0.21$          & 74.37 & 54.64$\pm$8.97 \\
59505.4 & 03111053015 & 67 & 1.1 & $3.4^{+0.7}_{-0.6}$    & $1.47\pm0.21$          & 63.11 & 57.64$\pm$8.53 \\
59525.9 & 03111053017 & 55 & 0.7 & $3.9^{+1.1}_{-0.6}$    & $1.55^{+0.20}_{-0.26}$ & 71.82 & 50.58$\pm$8.15 \\ \hline
59530.0	&	03111053018	&	84	&	0.8	&	\multirow{2}{*}{$5.8\pm0.6$}	&	\multirow{2}{*}{$1.50\pm0.11$}	&	\multirow{2}{*}{176.01}	&	\multirow{2}{*}{187.92$\pm$15.47}	\\ 
59530.2	&	03111053019	&	132	&	1.3	&		&		&		&		\\ \hline
59934.4 & 00080752003 & 62  & 0.9 & $4.1^{+0.9}_{-0.7}$    & $1.32\pm0.21$          & 48.14  & 58.95$\pm$8.77   \\
59965.8 & 00080752004 & 72  & 1.1 & $6.8^{+1.3}_{-1.1}$    & $1.20\pm0.20$          & 57.02  & 67.23$\pm$9.38   \\
60284.7 & 00016413001 & 199 & 3.0 & $3.6\pm0.4$            & $1.50\pm0.12$          & 118.90 & 168.55$\pm$15.04 \\
60285.6 & 00016413002 & 139 & 2.0 & $4.5\pm0.6$            & $1.31\pm0.14$          & 100.24 & 120.22$\pm$12.84 \\
60286.5 & 00016413003 & 175 & 2.8 & $3.6^{+0.5}_{-0.4}$    & $1.38\pm0.13$          & 154.01 & 146.31$\pm$14.04 \\
60287.2 & 00016413004 & 174 & 2.6 & $4.0\pm0.5$            & $1.31\pm0.13$          & 134.68 & 150.08$\pm$14.13 \\
60288.5 & 00016413005 & 202 & 2.8 & $5.4\pm0.6$            & $1.46\pm0.12$          & 126.58 & 167.32$\pm$15.28 \\
60289.1 & 00016413006 & 156 & 2.8 & $5.0^{+0.7}_{-0.6}$    & $1.08\pm0.14$          & 103.74 & 141.20$\pm$13.64 \\
60290.1 & 00016413007 & 156 & 2.4 & $3.7^{+0.5}_{-0.4}$    & $1.42\pm0.13$          & 98.43  & 135.02$\pm$13.44 \\
60294.4 & 00016413008 & 294 & 4.9 & $4.4\pm0.4$            & $1.29\pm0.10$          & 168.90 & 229.57$\pm$18.13 \\
60298.3 & 00016413009 & 84  & 2.0 & $3.5^{+0.7}_{-0.6}$    & $0.96^{+0.18}_{-0.19}$ & 72.38  & 76.15$\pm$10.19  \\
60302.1 & 00016413010 & 353 & 5.6 & $3.88^{+0.31}_{-0.29}$ & $1.24\pm0.09$          & 215.48 & 258.94$\pm$19.59 \\
60306.1 & 00016413011 & 235 & 5.3 & $2.60^{+0.25}_{-0.22}$ & $1.50\pm0.11$          & 143.20 & 181.87$\pm$16.22 \\
60310.0 & 00016413012 & 342 & 6.3 & $3.05^{+0.24}_{-0.22}$ & $1.46\pm0.09$          & 155.48 & 242.16$\pm$19.29 \\
60314.1 & 00016413013 & 200 & 3.5 & $3.5\pm0.4$            & $1.28\pm0.12$          & 134.41 & 157.03$\pm$14.80 \\
60318.1 & 00016413014 & 280 & 5.5 & $3.7\pm0.4$            & $1.22\pm0.10$          & 192.59 & 214.93$\pm$17.34 \\
60340.0 & 00016413015 & 153 & 2.7 & $3.4^{+0.5}_{-0.4}$    & $1.41\pm0.13$          & 112.10 & 125.51$\pm$13.01 \\
60347.2 & 00016413016 & 172 & 3.3 & $3.6^{+0.5}_{-0.4}$    & $1.40\pm0.13$          & 96.82  & 152.99$\pm$14.29 \\
60354.1 & 00016413017 & 131 & 2.9 & $3.7^{+0.6}_{-0.5}$    & $1.14\pm0.15$          & 89.53  & 123.92$\pm$12.79 \\
60361.0 & 00016413018 & 40  & 1.0 & $3.2^{+0.9}_{-0.7}$    & $1.06\pm0.26$          & 49.92  & 35.36$\pm$7.07   \\
60364.5 & 00016413019 & 111 & 2.3 & $3.4^{+0.6}_{-0.5}$    & $1.15\pm0.16$          & 107.70 & 99.21$\pm$11.58 \\
 \hline
\end{tabular}
\end{table*}

\bibliography{bibliography}{}

\begin{thebibliography}{}
\expandafter\ifx\csname natexlab\endcsname\relax\def\natexlab#1{#1}\fi
\providecommand{\url}[1]{\href{#1}{#1}}
\providecommand{\dodoi}[1]{doi:~\href{http://doi.org/#1}{\nolinkurl{#1}}}
\providecommand{\doeprint}[1]{\href{http://ascl.net/#1}{\nolinkurl{http://ascl.net/#1}}}
\providecommand{\doarXiv}[1]{\href{https://arxiv.org/abs/#1}{\nolinkurl{https://arxiv.org/abs/#1}}}

\bibitem[{{Abdo} {et~al.}(2010){Abdo}, {Ackermann}, {Ajello}, {Allafort}, {Antolini}, {Atwood}, {Axelsson}, {Baldini}, {Ballet}, {Barbiellini}, {Bastieri}, {Baughman}, {Bechtol}, {Bellazzini}, {Belli}, {Berenji}, {Bisello}, {Blandford}, {Bloom}, {Bonamente}, {Bonnell}, {Borgland}, {Bouvier}, {Bregeon}, {Brez}, {Brigida}, {Bruel}, {Burnett}, {Busetto}, {Buson}, {Caliandro}, {Cameron}, {Campana}, {Canadas}, {Caraveo}, {Carrigan}, {Casandjian}, {Cavazzuti}, {Ceccanti}, {Cecchi}, {{\c{C}}elik}, {Charles}, {Chekhtman}, {Cheung}, {Chiang}, {Cillis}, {Ciprini}, {Claus}, {Cohen-Tanugi}, {Conrad}, {Corbet}, {Davis}, {DeKlotz}, {den Hartog}, {Dermer}, {de Angelis}, {de Luca}, {de Palma}, {Digel}, {Dormody}, {Silva}, {Drell}, {Dubois}, {Dumora}, {Fabiani}, {Farnier}, {Favuzzi}, {Fegan}, {Ferrara}, {Focke}, {Fortin}, {Frailis}, {Fukazawa}, {Funk}, {Fusco}, {Gargano}, {Gasparrini}, {Gehrels}, {Germani}, {Giavitto}, {Giebels}, {Giglietto}, {Giommi}, {Giordano}, {Giroletti}, {Glanzman}, {Godfrey}, {Grenier}, {Grondin},
  {Grove}, {Guillemot}, {Guiriec}, {Gustafsson}, {Hadasch}, {Hanabata}, {Harding}, {Hayashida}, {Hays}, {Healey}, {Hill}, {Horan}, {Hughes}, {Iafrate}, {J{\'o}hannesson}, {Johnson}, {Johnson}, {Johnson}, {Johnson}, {Kamae}, {Katagiri}, {Kataoka}, {Kawai}, {Kerr}, {Kn{\"o}dlseder}, {Kocevski}, {Kuss}, {Lande}, {Landriu}, {Latronico}, {Lee}, {Lemoine-Goumard}, {Lionetto}, {Llena Garde}, {Longo}, {Loparco}, {Lott}, {Lovellette}, {Lubrano}, {Madejski}, {Makeev}, {Marangelli}, {Marelli}, {Massaro}, {Mazziotta}, {McConville}, {McEnery}, {Michelson}, {Minuti}, {Mitthumsiri}, {Mizuno}, {Moiseev}, {Mongelli}, {Monte}, {Monzani}, {Moretti}, {Morselli}, {Moskalenko}, {Murgia}, {Nakajima}, {Nakamori}, {Naumann-Godo}, {Nolan}, {Norris}, {Nuss}, {Ohno}, {Ohsugi}, {Omodei}, {Orlando}, {Ormes}, {Ozaki}, {Paccagnella}, {Paneque}, {Panetta}, {Parent}, {Pelassa}, {Pepe}, {Pesce-Rollins}, {Pinchera}, {Piron}, {Porter}, {Poupard}, {Rain{\`o}}, {Rando}, {Ray}, {Razzano}, {Razzaque}, {Rea}, {Reimer}, {Reimer}, {Reposeur}, {Ripken},
  {Ritz}, {Rochester}, {Rodriguez}, {Romani}, {Roth}, {Sadrozinski}, {Salvetti}, {Sanchez}, {Sander}, {Saz Parkinson}, {Scargle}, {Schalk}, {Scolieri}, {Sgr{\`o}}, {Shaw}, {Siskind}, {Smith}, {Smith}, \& {Spandre}}]{1fgl_cat}
{Abdo}, A.~A., {Ackermann}, M., {Ajello}, M., {et~al.} 2010, \apjs, 188, 405, \dodoi{10.1088/0067-0049/188/2/405}

\bibitem[{{Abdollahi} {et~al.}(2020){Abdollahi}, {Acero}, {Ackermann}, {Ajello}, {Atwood}, {Axelsson}, {Baldini}, {Ballet}, {Barbiellini}, {Bastieri}, {Becerra Gonzalez}, {Bellazzini}, {Berretta}, {Bissaldi}, {Blandford}, {Bloom}, {Bonino}, {Bottacini}, {Brandt}, {Bregeon}, {Bruel}, {Buehler}, {Burnett}, {Buson}, {Cameron}, {Caputo}, {Caraveo}, {Casandjian}, {Castro}, {Cavazzuti}, {Charles}, {Chaty}, {Chen}, {Cheung}, {Chiaro}, {Ciprini}, {Cohen-Tanugi}, {Cominsky}, {Coronado-Bl{\'a}zquez}, {Costantin}, {Cuoco}, {Cutini}, {D'Ammando}, {DeKlotz}, {de la Torre Luque}, {de Palma}, {Desai}, {Digel}, {Di Lalla}, {Di Mauro}, {Di Venere}, {Dom{\'\i}nguez}, {Dumora}, {Fana Dirirsa}, {Fegan}, {Ferrara}, {Franckowiak}, {Fukazawa}, {Funk}, {Fusco}, {Gargano}, {Gasparrini}, {Giglietto}, {Giommi}, {Giordano}, {Giroletti}, {Glanzman}, {Green}, {Grenier}, {Griffin}, {Grondin}, {Grove}, {Guiriec}, {Harding}, {Hayashi}, {Hays}, {Hewitt}, {Horan}, {J{\'o}hannesson}, {Johnson}, {Kamae}, {Kerr}, {Kocevski}, {Kovac'evic'},
  {Kuss}, {Landriu}, {Larsson}, {Latronico}, {Lemoine-Goumard}, {Li}, {Liodakis}, {Longo}, {Loparco}, {Lott}, {Lovellette}, {Lubrano}, {Madejski}, {Maldera}, {Malyshev}, {Manfreda}, {Marchesini}, {Marcotulli}, {Mart{\'\i}-Devesa}, {Martin}, {Massaro}, {Mazziotta}, {McEnery}, {Mereu}, {Meyer}, {Michelson}, {Mirabal}, {Mizuno}, {Monzani}, {Morselli}, {Moskalenko}, {Negro}, {Nuss}, {Ojha}, {Omodei}, {Orienti}, {Orlando}, {Ormes}, {Palatiello}, {Paliya}, {Paneque}, {Pei}, {Pe{\~n}a-Herazo}, {Perkins}, {Persic}, {Pesce-Rollins}, {Petrosian}, {Petrov}, {Piron}, {Poon}, {Porter}, {Principe}, {Rain{\`o}}, {Rando}, {Razzano}, {Razzaque}, {Reimer}, {Reimer}, {Remy}, {Reposeur}, {Romani}, {Saz Parkinson}, {Schinzel}, {Serini}, {Sgr{\`o}}, {Siskind}, {Smith}, {Spandre}, {Spinelli}, {Strong}, {Suson}, {Tajima}, {Takahashi}, {Tak}, {Thayer}, {Thompson}, {Tibaldo}, {Torres}, {Torresi}, {Valverde}, {Van Klaveren}, {van Zyl}, {Wood}, {Yassine}, \& {Zaharijas}}]{4fgl}
{Abdollahi}, S., {Acero}, F., {Ackermann}, M., {et~al.} 2020, \apjs, 247, 33, \dodoi{10.3847/1538-4365/ab6bcb}

\bibitem[{{Ackermann} {et~al.}(2017){Ackermann}, {Ajello}, {Baldini}, {Ballet}, {Barbiellini}, {Bastieri}, {Becerra Gonzalez}, {Bellazzini}, {Bissaldi}, {Blandford}, {Bloom}, {Bonino}, {Bottacini}, {Bregeon}, {Bruel}, {Buehler}, {Buson}, {Cameron}, {Caragiulo}, {Caraveo}, {Cavazzuti}, {Cecchi}, {Cheung}, {Chiang}, {Chiaro}, {Ciprini}, {Conrad}, {Costantin}, {Costanza}, {Cutini}, {D'Ammando}, {de Palma}, {Desiante}, {Digel}, {Di Lalla}, {Di Mauro}, {Di Venere}, {Dom{\'\i}nguez}, {Drell}, {Favuzzi}, {Fegan}, {Ferrara}, {Finke}, {Focke}, {Fukazawa}, {Funk}, {Fusco}, {Gargano}, {Gasparrini}, {Giglietto}, {Giordano}, {Giroletti}, {Green}, {Grenier}, {Guillemot}, {Guiriec}, {Hartmann}, {Hays}, {Horan}, {Jogler}, {J{\'o}hannesson}, {Johnson}, {Kuss}, {La Mura}, {Larsson}, {Latronico}, {Li}, {Longo}, {Loparco}, {Lovellette}, {Lubrano}, {Magill}, {Maldera}, {Manfreda}, {Marcotulli}, {Mazziotta}, {Michelson}, {Mirabal}, {Mitthumsiri}, {Mizuno}, {Monzani}, {Morselli}, {Moskalenko}, {Negro}, {Nuss}, {Ohsugi}, {Ojha},
  {Omodei}, {Orienti}, {Orlando}, {Ormes}, {Paliya}, {Paneque}, {Perkins}, {Persic}, {Pesce-Rollins}, {Piron}, {Porter}, {Principe}, {Rain{\`o}}, {Rando}, {Rani}, {Razzano}, {Razzaque}, {Reimer}, {Reimer}, {Romani}, {Sgr{\`o}}, {Simone}, {Siskind}, {Spada}, {Spandre}, {Spinelli}, {Stalin}, {Stawarz}, {Suson}, {Takahashi}, {Tanaka}, {Thayer}, {Thompson}, {Torres}, {Torresi}, {Tosti}, {Troja}, {Vianello}, \& {Wood}}]{ackermann2017}
{Ackermann}, M., {Ajello}, M., {Baldini}, L., {et~al.} 2017, \apjl, 837, L5, \dodoi{10.3847/2041-8213/aa5fff}

\bibitem[{Agudo {et~al.}(2012)Agudo, Molina, G\'omez, Marscher, Jorstad, \& Heidt}]{agudo2012}
Agudo, I., Molina, S., G\'omez, J.~L., {et~al.} 2012, International Journal of Modern Physics: Conference Series, 08, 299, \dodoi{10.1142/S2010194512004746}

\bibitem[{{Ajello} {et~al.}(2022){Ajello}, {Baldini}, {Ballet}, {Bastieri}, {Becerra Gonzalez}, {Bellazzini}, {Berretta}, {Bissaldi}, {Bonino}, {Brill}, {Bruel}, {Buson}, {Caputo}, {Caraveo}, {Cheung}, {Chiaro}, {Cibrario}, {Ciprini}, {Crnogorcevic}, {Cutini}, {D'Ammando}, {De Gaetano}, {Di Lalla}, {Di Venere}, {Dom{\'\i}nguez}, {Ramazani}, {Ferrara}, {Fiori}, {Fukazawa}, {Funk}, {Fusco}, {Gammaldi}, {Gargano}, {Garrappa}, {Gasparrini}, {Giglietto}, {Giordano}, {Giroletti}, {Green}, {Grenier}, {Guiriec}, {Horan}, {Hou}, {Kayanoki}, {Kuss}, {Larsson}, {Latronico}, {Lewis}, {Li}, {Liodakis}, {Longo}, {Loparco}, {Lott}, {Lovellette}, {Lubrano}, {Madejski}, {Maldera}, {Manfreda}, {Mart{\'\i}-Devesa}, {Mazziotta}, {Mereu}, {Michelson}, {Mirabal}, {Mitthumsiri}, {Mizuno}, {Monzani}, {Morselli}, {Moskalenko}, {Negro}, {Ojha}, {Orienti}, {Orlando}, {Ormes}, {Pei}, {Pe{\~n}a-Herazo}, {Persic}, {Pesce-Rollins}, {Petrosian}, {Pillera}, {Poon}, {Porter}, {Principe}, {Rain{\`o}}, {Rando}, {Rani}, {Razzano}, {Razzaque},
  {Reimer}, {Reimer}, {Scotton}, {Serini}, {Sgr{\`o}}, {Siskind}, {Spandre}, {Spinelli}, {Suson}, {Tajima}, {Torres}, {Valverde}, {Yassin}, \& {Zaharijas}}]{4lac-dr3}
{Ajello}, M., {Baldini}, L., {Ballet}, J., {et~al.} 2022, \apjs, 263, 24, \dodoi{10.3847/1538-4365/ac9523}

\bibitem[{{Angioni} {et~al.}(2019){Angioni}, {Nesci}, {Finke}, {Buson}, \& {Ciprini}}]{angioni2019}
{Angioni}, R., {Nesci}, R., {Finke}, J.~D., {Buson}, S., \& {Ciprini}, S. 2019, \aap, 627, A140, \dodoi{10.1051/0004-6361/201935461}

\bibitem[{{Ba{\~n}ados} {et~al.}(2024){Ba{\~n}ados}, {Momjian}, {Connor}, {Belladitta}, {Decarli}, {Mazzucchelli}, {Venemans}, {Walter}, {Wang}, {Xie}, {Barth}, {Eilers}, {Fan}, {Khusanova}, {Schindler}, {Stern}, {Yang}, {Andika}, {Carilli}, {Farina}, {Fabian}, {Hennawi}, {Pensabene}, \& {Rojas-Ruiz}}]{banados2024}
{Ba{\~n}ados}, E., {Momjian}, E., {Connor}, T., {et~al.} 2024, Nature Astronomy, \dodoi{10.1038/s41550-024-02431-4}

\bibitem[{{Ballet} {et~al.}(2023){Ballet}, {Bruel}, {Burnett}, {Lott}, \& {The Fermi-LAT collaboration}}]{4fgl-dr4}
{Ballet}, J., {Bruel}, P., {Burnett}, T.~H., {Lott}, B., \& {The Fermi-LAT collaboration}. 2023, arXiv e-prints, arXiv:2307.12546, \dodoi{10.48550/arXiv.2307.12546}

\bibitem[{{Begelman} \& {Sikora}(1987)}]{begelman1987}
{Begelman}, M.~C., \& {Sikora}, M. 1987, \apj, 322, 650, \dodoi{10.1086/165760}

\bibitem[{{Begelman} {et~al.}(2006){Begelman}, {Volonteri}, \& {Rees}}]{begelman2006}
{Begelman}, M.~C., {Volonteri}, M., \& {Rees}, M.~J. 2006, \mnras, 370, 289, \dodoi{10.1111/j.1365-2966.2006.10467.x}

\bibitem[{{Belladitta} {et~al.}(2020){Belladitta}, {Moretti}, {Caccianiga}, {Spingola}, {Severgnini}, {Della Ceca}, {Ghisellini}, {Dallacasa}, {Sbarrato}, {Cicone}, {Cassar{\`a}}, \& {Pedani}}]{belladitta2020}
{Belladitta}, S., {Moretti}, A., {Caccianiga}, A., {et~al.} 2020, \aap, 635, L7, \dodoi{10.1051/0004-6361/201937395}

\bibitem[{{Benke} {et~al.}(2024){Benke}, {Gokus}, {Lisakov}, {Gurvits}, {Eppel}, {He{\ss}d{\"o}rfer}, {Kadler}, {Kovalev}, {Ros}, \& {R{\"o}sch}}]{benke2024}
{Benke}, P., {Gokus}, A., {Lisakov}, M., {et~al.} 2024, \aap, 689, A43, \dodoi{10.1051/0004-6361/202450153}

\bibitem[{{Bogd{\'a}n} {et~al.}(2024){Bogd{\'a}n}, {Goulding}, {Natarajan}, {Kov{\'a}cs}, {Tremblay}, {Chadayammuri}, {Volonteri}, {Kraft}, {Forman}, {Jones}, {Churazov}, \& {Zhuravleva}}]{bogdan2024}
{Bogd{\'a}n}, {\'A}., {Goulding}, A.~D., {Natarajan}, P., {et~al.} 2024, Nature Astronomy, 8, 126, \dodoi{10.1038/s41550-023-02111-9}

\bibitem[{{Boller} {et~al.}(2000){Boller}, {Fabian}, {Brandt}, \& {Freyberg}}]{Boller2000}
{Boller}, T., {Fabian}, A.~C., {Brandt}, W.~N., \& {Freyberg}, M.~J. 2000, \mnras, 315, L23, \dodoi{10.1046/j.1365-8711.2000.03635.x}

\bibitem[{{B{\"o}ttcher} {et~al.}(2013){B{\"o}ttcher}, {Reimer}, {Sweeney}, \& {Prakash}}]{Boettcher13}
{B{\"o}ttcher}, M., {Reimer}, A., {Sweeney}, K., \& {Prakash}, A. 2013, \apj, 768, 54, \dodoi{10.1088/0004-637X/768/1/54}

\bibitem[{{Bromm} \& {Loeb}(2003)}]{Bromm2003}
{Bromm}, V., \& {Loeb}, A. 2003, \apj, 596, 34, \dodoi{10.1086/377529}

\bibitem[{{Burke} {et~al.}(2024){Burke}, {Liu}, \& {Shen}}]{burke2024}
{Burke}, C.~J., {Liu}, X., \& {Shen}, Y. 2024, \mnras, 527, 5356, \dodoi{10.1093/mnras/stad3592}

\bibitem[{{Cammelli} {et~al.}(2025){Cammelli}, {Monaco}, {Tan}, {Singh}, {Fontanot}, {De Lucia}, {Hirschmann}, \& {Xie}}]{cammelli2025}
{Cammelli}, V., {Monaco}, P., {Tan}, J.~C., {et~al.} 2025, \mnras, 536, 851, \dodoi{10.1093/mnras/stae2663}

\bibitem[{{Cash}(1979)}]{cash1979}
{Cash}, W. 1979, \apj, 228, 939, \dodoi{10.1086/156922}

\bibitem[{{Celotti} {et~al.}(2007){Celotti}, {Ghisellini}, \& {Fabian}}]{celotti2007}
{Celotti}, A., {Ghisellini}, G., \& {Fabian}, A.~C. 2007, \mnras, 375, 417, \dodoi{10.1111/j.1365-2966.2006.11289.x}

\bibitem[{{Cheung} {et~al.}(2012){Cheung}, {Stawarz}, {Siemiginowska}, {Gobeille}, {Wardle}, {Harris}, \& {Schwartz}}]{Cheung2012}
{Cheung}, C.~C., {Stawarz}, {\L}., {Siemiginowska}, A., {et~al.} 2012, \apjl, 756, L20, \dodoi{10.1088/2041-8205/756/1/L20}

\bibitem[{{Diana} {et~al.}(2022){Diana}, {Caccianiga}, {Ighina}, {Belladitta}, {Moretti}, \& {Della Ceca}}]{diana2022}
{Diana}, A., {Caccianiga}, A., {Ighina}, L., {et~al.} 2022, \mnras, 511, 5436, \dodoi{10.1093/mnras/stac364}

\bibitem[{{Dom{\'\i}nguez} {et~al.}(2024){Dom{\'\i}nguez}, {{\O}stergaard Kirkeberg}, {Wojtak}, {Saldana-Lopez}, {Desai}, {Primack}, {Finke}, {Ajello}, {P{\'e}rez-Gonz{\'a}lez}, {Paliya}, \& {Hartmann}}]{dominguez2024}
{Dom{\'\i}nguez}, A., {{\O}stergaard Kirkeberg}, P., {Wojtak}, R., {et~al.} 2024, \mnras, 527, 4632, \dodoi{10.1093/mnras/stad3425}

\bibitem[{{Eppel} {et~al.}(2024){Eppel}, {Kadler}, {He{\ss}d{\"o}rfer}, {Benke}, {Debbrecht}, {Eich}, {Gokus}, {H{\"a}mmerich}, {Kirchner}, {Paraschos}, {R{\"o}sch}, {Schulga}, {Sinapius}, {Weber}, {Bach}, {Dorner}, {Edwards}, {Giroletti}, {Kraus}, {Hervet}, {Koyama}, {Krichbaum}, {Mannheim}, {Ros}, {Zacharias}, \& {Zensus}}]{eppel2024}
{Eppel}, F., {Kadler}, M., {He{\ss}d{\"o}rfer}, J., {et~al.} 2024, \aap, 684, A11, \dodoi{10.1051/0004-6361/202348262}

\bibitem[{{Escudero Pedrosa} {et~al.}(2024{\natexlab{a}}){Escudero Pedrosa}, {Morcuende Parrilla}, \& {Otero-Santos}}]{escudero2024b}
{Escudero Pedrosa}, J., {Morcuende Parrilla}, D., \& {Otero-Santos}, J. 2024{\natexlab{a}}, {IOP4}, v1.4.1,  Zenodo, \dodoi{10.5281/zenodo.10222722}

\bibitem[{{Escudero Pedrosa} {et~al.}(2024{\natexlab{b}}){Escudero Pedrosa}, {Agudo}, {Morcuende}, {Otero-Santos}, {Bonnoli}, {Piirola}, {Husillos}, {Bernardos}, {L{\'o}pez-Coto}, {Sota}, {Casanova}, {Aceituno}, \& {Santos-Sanz}}]{escudero2024a}
{Escudero Pedrosa}, J., {Agudo}, I., {Morcuende}, D., {et~al.} 2024{\natexlab{b}}, \aj, 168, 84, \dodoi{10.3847/1538-3881/ad5a80}

\bibitem[{{Fabian} {et~al.}(1997){Fabian}, {Brandt}, {McMahon}, \& {Hook}}]{fabian1997}
{Fabian}, A.~C., {Brandt}, W.~N., {McMahon}, R.~G., \& {Hook}, I.~M. 1997, \mnras, 291, L5, \dodoi{10.1093/mnras/291.1.L5}

\bibitem[{{Fabian} {et~al.}(2001){Fabian}, {Celotti}, {Iwasawa}, \& {Ghisellini}}]{Fabian2001}
{Fabian}, A.~C., {Celotti}, A., {Iwasawa}, K., \& {Ghisellini}, G. 2001, \mnras, 324, 628, \dodoi{10.1046/j.1365-8711.2001.04348.x}

\bibitem[{{Fabian} {et~al.}(1999){Fabian}, {Celotti}, {Pooley}, {Iwasawa}, {Brandt}, {McMahon}, \& {Hoenig}}]{fabian1999}
{Fabian}, A.~C., {Celotti}, A., {Pooley}, G., {et~al.} 1999, \mnras, 308, L6, \dodoi{10.1046/j.1365-8711.1999.02910.x}

\bibitem[{{Fabian} {et~al.}(1998){Fabian}, {Iwasawa}, {McMahon}, {Celotti}, {Brandt}, \& {Hook}}]{fabian1998}
{Fabian}, A.~C., {Iwasawa}, K., {McMahon}, R.~G., {et~al.} 1998, \mnras, 295, L25, \dodoi{10.1046/j.1365-8711.1998.29511483.x}

\bibitem[{{Fan} {et~al.}(2017){Fan}, {Yang}, {Xiao}, {Lin}, {Constantin}, {Luo}, {Pei}, {Hao}, \& {Mao}}]{fan2017}
{Fan}, J.~H., {Yang}, J.~H., {Xiao}, H.~B., {et~al.} 2017, \apjl, 835, L38, \dodoi{10.3847/2041-8213/835/2/L38}

\bibitem[{{Fossati} {et~al.}(1998){Fossati}, {Maraschi}, {Celotti}, {Comastri}, \& {Ghisellini}}]{fossati1998}
{Fossati}, G., {Maraschi}, L., {Celotti}, A., {Comastri}, A., \& {Ghisellini}, G. 1998, \mnras, 299, 433, \dodoi{10.1046/j.1365-8711.1998.01828.x}

\bibitem[{{Ghisellini} {et~al.}(2013){Ghisellini}, {Haardt}, {Della Ceca}, {Volonteri}, \& {Sbarrato}}]{ghisellini2013}
{Ghisellini}, G., {Haardt}, F., {Della Ceca}, R., {Volonteri}, M., \& {Sbarrato}, T. 2013, \mnras, 432, 2818, \dodoi{10.1093/mnras/stt637}

\bibitem[{{Ghisellini} {et~al.}(2017){Ghisellini}, {Righi}, {Costamante}, \& {Tavecchio}}]{ghisellini2017}
{Ghisellini}, G., {Righi}, C., {Costamante}, L., \& {Tavecchio}, F. 2017, \mnras, 469, 255, \dodoi{10.1093/mnras/stx806}

\bibitem[{{Ghisellini} \& {Tavecchio}(2009)}]{ghisellini2009}
{Ghisellini}, G., \& {Tavecchio}, F. 2009, \mnras, 397, 985, \dodoi{10.1111/j.1365-2966.2009.15007.x}

\bibitem[{{Giommi} {et~al.}(2012){Giommi}, {Padovani}, {Polenta}, {Turriziani}, {D'Elia}, \& {Piranomonte}}]{giommi2012}
{Giommi}, P., {Padovani}, P., {Polenta}, G., {et~al.} 2012, \mnras, 420, 2899, \dodoi{10.1111/j.1365-2966.2011.20044.x}

\bibitem[{{Gokus} {et~al.}(2021){Gokus}, {Paliya}, {Wagner}, {Buson}, {D'Ammando}, {Edwards}, {Kadler}, {Meyer}, {Ojha}, {Stevens}, \& {Wilms}}]{gokus2021}
{Gokus}, A., {Paliya}, V.~S., {Wagner}, S.~M., {et~al.} 2021, \aap, 649, A77, \dodoi{10.1051/0004-6361/202039378}

\bibitem[{{Gokus} {et~al.}(2024){Gokus}, {B{\"o}ttcher}, {Errando}, {Kreter}, {He{\ss}d{\"o}rfer}, {Eppel}, {Kadler}, {Smith}, {Benke}, {Gurvits}, {Kraus}, {Lisakov}, {McBride}, {Ros}, {R{\"o}sch}, \& {Wilms}}]{Gokus2024}
{Gokus}, A., {B{\"o}ttcher}, M., {Errando}, M., {et~al.} 2024, \apj, 974, 38, \dodoi{10.3847/1538-4357/ad6a4e}

\bibitem[{{Hayashida} {et~al.}(2015){Hayashida}, {Nalewajko}, {Madejski}, {Sikora}, {Itoh}, {Ajello}, {Blandford}, {Buson}, {Chiang}, {Fukazawa}, {Furniss}, {Urry}, {Hasan}, {Harrison}, {Alexander}, {Balokovi{\'c}}, {Barret}, {Boggs}, {Christensen}, {Craig}, {Forster}, {Giommi}, {Grefenstette}, {Hailey}, {Hornstrup}, {Kitaguchi}, {Koglin}, {Madsen}, {Mao}, {Miyasaka}, {Mori}, {Perri}, {Pivovaroff}, {Puccetti}, {Rana}, {Stern}, {Tagliaferri}, {Westergaard}, {Zhang}, {Zoglauer}, {Gurwell}, {Uemura}, {Akitaya}, {Kawabata}, {Kawaguchi}, {Kanda}, {Moritani}, {Takaki}, {Ui}, {Yoshida}, {Agarwal}, \& {Gupta}}]{hayashida2015}
{Hayashida}, M., {Nalewajko}, K., {Madejski}, G.~M., {et~al.} 2015, \apj, 807, 79, \dodoi{10.1088/0004-637X/807/1/79}

\bibitem[{{HI4PI Collaboration} {et~al.}(2016){HI4PI Collaboration}, {Ben Bekhti}, {Fl{\"o}er}, {Keller}, {Kerp}, {Lenz}, {Winkel}, {Bailin}, {Calabretta}, {Dedes}, {Ford}, {Gibson}, {Haud}, {Janowiecki}, {Kalberla}, {Lockman}, {McClure-Griffiths}, {Murphy}, {Nakanishi}, {Pisano}, \& {Staveley-Smith}}]{HI4PI}
{HI4PI Collaboration}, {Ben Bekhti}, N., {Fl{\"o}er}, L., {et~al.} 2016, \aap, 594, A116, \dodoi{10.1051/0004-6361/201629178}

\bibitem[{{Hook} \& {McMahon}(1998)}]{Hook1998}
{Hook}, I.~M., \& {McMahon}, R.~G. 1998, \mnras, 294, L7, \dodoi{10.1046/j.1365-8711.1998.01368.x}

\bibitem[{{Houck} \& {Denicola}(2000)}]{isis}
{Houck}, J.~C., \& {Denicola}, L.~A. 2000, in Astronomical Data Analysis Software and Systems IX, ed. N.~{Manset}, C.~{Veillet}, \& D.~{Crabtree}, ASP Conf.\ Ser. No. 216 (San Francisco: Astron.\ Soc.\ Pacific), 591

\bibitem[{{Ighina} {et~al.}(2024){Ighina}, {Caccianiga}, {Moretti}, {Broderick}, {Leung}, {L{\'o}pez-S{\'a}nchez}, {Rigamonti}, {Seymour}, {An}, {Belladitta}, {Bisogni}, {Della Ceca}, {Drouart}, {Gargiulo}, \& {Liu}}]{ighina2024}
{Ighina}, L., {Caccianiga}, A., {Moretti}, A., {et~al.} 2024, \aap, 692, A241, \dodoi{10.1051/0004-6361/202451376}

\bibitem[{{Inayoshi} {et~al.}(2020){Inayoshi}, {Visbal}, \& {Haiman}}]{Inayoshi2020}
{Inayoshi}, K., {Visbal}, E., \& {Haiman}, Z. 2020, \araa, 58, 27, \dodoi{10.1146/annurev-astro-120419-014455}

\bibitem[{{Kaastra} \& {Bleeker}(2016)}]{optimalbinning}
{Kaastra}, J.~S., \& {Bleeker}, J.~A.~M. 2016, \aap, 587, A151, \dodoi{10.1051/0004-6361/201527395}

\bibitem[{{Kamaram} {et~al.}(2023){Kamaram}, {Prince}, {Pramanick}, \& {Bose}}]{kamaram2023}
{Kamaram}, S.~R., {Prince}, R., {Pramanick}, S., \& {Bose}, D. 2023, \mnras, 520, 2024, \dodoi{10.1093/mnras/stad167}

\bibitem[{{Keenan} {et~al.}(2021){Keenan}, {Meyer}, {Georganopoulos}, {Reddy}, \& {French}}]{keenan2021}
{Keenan}, M., {Meyer}, E.~T., {Georganopoulos}, M., {Reddy}, K., \& {French}, O.~J. 2021, \mnras, 505, 4726, \dodoi{10.1093/mnras/stab1182}

\bibitem[{{Kreter} {et~al.}(2020){Kreter}, {Gokus}, {Krauss}, {Kadler}, {Ojha}, {Buson}, {Wilms}, \& {B{\"o}ttcher}}]{kreter2020}
{Kreter}, M., {Gokus}, A., {Krauss}, F., {et~al.} 2020, \apj, 903, 128, \dodoi{10.3847/1538-4357/abb8da}

\bibitem[{{Liao} {et~al.}(2019){Liao}, {Dou}, {Jiang}, {Wang}, {Fan}, \& {Wang}}]{Liao2019}
{Liao}, N.-H., {Dou}, L.-M., {Jiang}, N., {et~al.} 2019, \apjl, 879, L9, \dodoi{10.3847/2041-8213/ab2893}

\bibitem[{{Liao} {et~al.}(2018){Liao}, {Li}, \& {Fan}}]{liao2018}
{Liao}, N.-H., {Li}, S., \& {Fan}, Y.-Z. 2018, \apjl, 865, L17, \dodoi{10.3847/2041-8213/aae20d}

\bibitem[{{Lodato} \& {Natarajan}(2006)}]{lodato2006}
{Lodato}, G., \& {Natarajan}, P. 2006, \mnras, 371, 1813, \dodoi{10.1111/j.1365-2966.2006.10801.x}

\bibitem[{{Lyke} {et~al.}(2020){Lyke}, {Higley}, {McLane}, {Schurhammer}, {Myers}, {Ross}, {Dawson}, {Chabanier}, {Martini}, {Busca}, {Mas des Bourboux}, {Salvato}, {Streblyanska}, {Zarrouk}, {Burtin}, {Anderson}, {Bautista}, {Bizyaev}, {Brandt}, {Brinkmann}, {Brownstein}, {Comparat}, {Green}, {de la Macorra}, {Mu{\~n}oz Guti{\'e}rrez}, {Hou}, {Newman}, {Palanque-Delabrouille}, {P{\^a}ris}, {Percival}, {Petitjean}, {Rich}, {Rossi}, {Schneider}, {Smith}, {Vivek}, \& {Weaver}}]{lyke2020}
{Lyke}, B.~W., {Higley}, A.~N., {McLane}, J.~N., {et~al.} 2020, \apjs, 250, 8, \dodoi{10.3847/1538-4365/aba623}

\bibitem[{{Madau} \& {Rees}(2001)}]{madau2001}
{Madau}, P., \& {Rees}, M.~J. 2001, \apjl, 551, L27, \dodoi{10.1086/319848}

\bibitem[{{Marcotulli} {et~al.}(2020){Marcotulli}, {Paliya}, {Ajello}, {Kaur}, {Marchesi}, {Rajagopal}, {Hartmann}, {Gasparrini}, {Ojha}, \& {Madejski}}]{Marcotulli2020}
{Marcotulli}, L., {Paliya}, V., {Ajello}, M., {et~al.} 2020, \apj, 889, 164, \dodoi{10.3847/1538-4357/ab65f5}

\bibitem[{{Marcotulli} {et~al.}(2022){Marcotulli}, {Ajello}, {Urry}, {Paliya}, {Koss}, {Oh}, {Madejski}, {Ueda}, {Balokovi{\'c}}, {Trakhtenbrot}, {Ricci}, {Ricci}, {Stern}, {Harrison}, {Powell}, \& {BASS Collaboration}}]{marcotulli2022}
{Marcotulli}, L., {Ajello}, M., {Urry}, C.~M., {et~al.} 2022, \apj, 940, 77, \dodoi{10.3847/1538-4357/ac937f}

\bibitem[{{Marcotulli} {et~al.}(2025){Marcotulli}, {Connor}, {Ba{\~n}ados}, {Boorman}, {Migliori}, {Grefenstette}, {Momjian}, {Siemiginowska}, {Stern}, {Belladitta}, {Cheung}, {Fabian}, {Khusanova}, {Mazzucchelli}, {Rojas-Ruiz}, \& {Urry}}]{marcotulli2025}
{Marcotulli}, L., {Connor}, T., {Ba{\~n}ados}, E., {et~al.} 2025, arXiv e-prints, arXiv:2501.07637.
\newblock \doarXiv{2501.07637}

\bibitem[{{Massaro} {et~al.}(2009){Massaro}, {Giommi}, {Leto}, {Marchegiani}, {Maselli}, {Perri}, {Piranomonte}, \& {Sclavi}}]{bzcat_1stedition}
{Massaro}, E., {Giommi}, P., {Leto}, C., {et~al.} 2009, \aap, 495, 691, \dodoi{10.1051/0004-6361:200810161}

\bibitem[{{Massaro} {et~al.}(2015){Massaro}, {Maselli}, {Leto}, {Marchegiani}, {Perri}, {Giommi}, \& {Piranomonte}}]{bzcat_5thedition}
{Massaro}, E., {Maselli}, A., {Leto}, C., {et~al.} 2015, \apss, 357, 75, \dodoi{10.1007/s10509-015-2254-2}

\bibitem[{{Mattox} {et~al.}(1996){Mattox}, {Bertsch}, {Chiang}, {Dingus}, {Digel}, {Esposito}, {Fierro}, {Hartman}, {Hunter}, {Kanbach}, {Kniffen}, {Lin}, {Macomb}, {Mayer-Hasselwander}, {Michelson}, {von Montigny}, {Mukherjee}, {Nolan}, {Ramanamurthy}, {Schneid}, {Sreekumar}, {Thompson}, \& {Willis}}]{mattox}
{Mattox}, J.~R., {Bertsch}, D.~L., {Chiang}, J., {et~al.} 1996, \apj, 461, 396, \dodoi{10.1086/177068}

\bibitem[{{McKinney} {et~al.}(2014){McKinney}, {Tchekhovskoy}, {Sadowski}, \& {Narayan}}]{mckinney2014}
{McKinney}, J.~C., {Tchekhovskoy}, A., {Sadowski}, A., \& {Narayan}, R. 2014, \mnras, 441, 3177, \dodoi{10.1093/mnras/stu762}

\bibitem[{{Medvedev} {et~al.}(2020){Medvedev}, {Sazonov}, {Gilfanov}, {Burenin}, {Khorunzhev}, {Meshcheryakov}, {Sunyaev}, {Bikmaev}, \& {Irtuganov}}]{medvedev2020}
{Medvedev}, P., {Sazonov}, S., {Gilfanov}, M., {et~al.} 2020, \mnras, 497, 1842, \dodoi{10.1093/mnras/staa2051}

\bibitem[{{Meyer} {et~al.}(2011){Meyer}, {Fossati}, {Georganopoulos}, \& {Lister}}]{meyer2011}
{Meyer}, E.~T., {Fossati}, G., {Georganopoulos}, M., \& {Lister}, M.~L. 2011, \apj, 740, 98, \dodoi{10.1088/0004-637X/740/2/98}

\bibitem[{{Meyer} {et~al.}(2019){Meyer}, {Scargle}, \& {Blandford}}]{Meyer2019}
{Meyer}, M., {Scargle}, J.~D., \& {Blandford}, R.~D. 2019, \apj, 877, 39, \dodoi{10.3847/1538-4357/ab1651}

\bibitem[{{Migliori} {et~al.}(2023){Migliori}, {Siemiginowska}, {Sobolewska}, {Cheung}, {Stawarz}, {Schwartz}, {Snios}, {Saxena}, \& {Kashyap}}]{migliori2023}
{Migliori}, G., {Siemiginowska}, A., {Sobolewska}, M., {et~al.} 2023, \mnras, 524, 1087, \dodoi{10.1093/mnras/stad1959}

\bibitem[{{Moretti} {et~al.}(2021){Moretti}, {Ghisellini}, {Caccianiga}, {Belladitta}, {Della Ceca}, {Ighina}, {Sbarrato}, {Severgnini}, \& {Spingola}}]{moretti2021}
{Moretti}, A., {Ghisellini}, G., {Caccianiga}, A., {et~al.} 2021, \apj, 920, 15, \dodoi{10.3847/1538-4357/ac167a}

\bibitem[{{Nalewajko} \& {Gupta}(2017)}]{2017A&A...606A..44N}
{Nalewajko}, K., \& {Gupta}, M. 2017, \aap, 606, A44, \dodoi{10.1051/0004-6361/201731329}

\bibitem[{{Nasa High Energy Astrophysics Science Archive Research Center (Heasarc)}(2014)}]{heasoft}
{Nasa High Energy Astrophysics Science Archive Research Center (Heasarc)}. 2014, {HEAsoft: Unified Release of FTOOLS and XANADU}, Astrophysics Source Code Library, record ascl:1408.004

\bibitem[{{Nieppola} {et~al.}(2008){Nieppola}, {Valtaoja}, {Tornikoski}, {Hovatta}, \& {Kotiranta}}]{nieppola2008}
{Nieppola}, E., {Valtaoja}, E., {Tornikoski}, M., {Hovatta}, T., \& {Kotiranta}, M. 2008, \aap, 488, 867, \dodoi{10.1051/0004-6361:200809716}

\bibitem[{{Pacciani} {et~al.}(2014){Pacciani}, {Tavecchio}, {Donnarumma}, {Stamerra}, {Carrasco}, {Recillas}, {Porras}, \& {Uemura}}]{pacciani2014}
{Pacciani}, L., {Tavecchio}, F., {Donnarumma}, I., {et~al.} 2014, \apj, 790, 45, \dodoi{10.1088/0004-637X/790/1/45}

\bibitem[{{Paliya} {et~al.}(2020){Paliya}, {Ajello}, {Cao}, {Giroletti}, {Kaur}, {Madejski}, {Lott}, \& {Hartmann}}]{Paliya2020}
{Paliya}, V.~S., {Ajello}, M., {Cao}, H.~M., {et~al.} 2020, \apj, 897, 177, \dodoi{10.3847/1538-4357/ab9c1a}

\bibitem[{{Paliya} {et~al.}(2021){Paliya}, {Dom{\'\i}nguez}, {Ajello}, {Olmo-Garc{\'\i}a}, \& {Hartmann}}]{paliya2021}
{Paliya}, V.~S., {Dom{\'\i}nguez}, A., {Ajello}, M., {Olmo-Garc{\'\i}a}, A., \& {Hartmann}, D. 2021, \apjs, 253, 46, \dodoi{10.3847/1538-4365/abe135}

\bibitem[{{Paliya} {et~al.}(2016){Paliya}, {Parker}, {Fabian}, \& {Stalin}}]{Paliya2016}
{Paliya}, V.~S., {Parker}, M.~L., {Fabian}, A.~C., \& {Stalin}, C.~S. 2016, \apj, 825, 74, \dodoi{10.3847/0004-637X/825/1/74}

\bibitem[{{Paliya} {et~al.}(2019){Paliya}, {Ajello}, {Ojha}, {Angioni}, {Cheung}, {Tanada}, {Pursimo}, {Galindo}, {Losada}, {Siltala}, {Djupvik}, {Marcotulli}, \& {Hartmann}}]{Paliya2019}
{Paliya}, V.~S., {Ajello}, M., {Ojha}, R., {et~al.} 2019, \apj, 871, 211, \dodoi{10.3847/1538-4357/aafa10}

\bibitem[{{Planck Collaboration} {et~al.}(2016){Planck Collaboration}, {Ade}, {Aghanim}, {Arnaud}, {Ashdown}, {Aumont}, {Baccigalupi}, {Banday}, {Barreiro}, {Bartlett}, {Bartolo}, {Battaner}, {Battye}, {Benabed}, {Beno{\^\i}t}, {Benoit-L{\'e}vy}, {Bernard}, {Bersanelli}, {Bielewicz}, {Bock}, {Bonaldi}, {Bonavera}, {Bond}, {Borrill}, {Bouchet}, {Boulanger}, {Bucher}, {Burigana}, {Butler}, {Calabrese}, {Cardoso}, {Catalano}, {Challinor}, {Chamballu}, {Chary}, {Chiang}, {Chluba}, {Christensen}, {Church}, {Clements}, {Colombi}, {Colombo}, {Combet}, {Coulais}, {Crill}, {Curto}, {Cuttaia}, {Danese}, {Davies}, {Davis}, {de Bernardis}, {de Rosa}, {de Zotti}, {Delabrouille}, {D{\'e}sert}, {Di Valentino}, {Dickinson}, {Diego}, {Dolag}, {Dole}, {Donzelli}, {Dor{\'e}}, {Douspis}, {Ducout}, {Dunkley}, {Dupac}, {Efstathiou}, {Elsner}, {En{\ss}lin}, {Eriksen}, {Farhang}, {Fergusson}, {Finelli}, {Forni}, {Frailis}, {Fraisse}, {Franceschi}, {Frejsel}, {Galeotta}, {Galli}, {Ganga}, {Gauthier}, {Gerbino}, {Ghosh}, {Giard},
  {Giraud-H{\'e}raud}, {Giusarma}, {Gjerl{\o}w}, {Gonz{\'a}lez-Nuevo}, {G{\'o}rski}, {Gratton}, {Gregorio}, {Gruppuso}, {Gudmundsson}, {Hamann}, {Hansen}, {Hanson}, {Harrison}, {Helou}, {Henrot-Versill{\'e}}, {Hern{\'a}ndez-Monteagudo}, {Herranz}, {Hildebrandt}, {Hivon}, {Hobson}, {Holmes}, {Hornstrup}, {Hovest}, {Huang}, {Huffenberger}, {Hurier}, {Jaffe}, {Jaffe}, {Jones}, {Juvela}, {Keih{\"a}nen}, {Keskitalo}, {Kisner}, {Kneissl}, {Knoche}, {Knox}, {Kunz}, {Kurki-Suonio}, {Lagache}, {L{\"a}hteenm{\"a}ki}, {Lamarre}, {Lasenby}, {Lattanzi}, {Lawrence}, {Leahy}, {Leonardi}, {Lesgourgues}, {Levrier}, {Lewis}, {Liguori}, {Lilje}, {Linden-V{\o}rnle}, {L{\'o}pez-Caniego}, {Lubin}, {Mac{\'\i}as-P{\'e}rez}, {Maggio}, {Maino}, {Mandolesi}, {Mangilli}, {Marchini}, {Maris}, {Martin}, {Martinelli}, {Mart{\'\i}nez-Gonz{\'a}lez}, {Masi}, {Matarrese}, {McGehee}, {Meinhold}, {Melchiorri}, {Melin}, {Mendes}, {Mennella}, {Migliaccio}, {Millea}, {Mitra}, {Miville-Desch{\^e}nes}, {Moneti}, {Montier}, {Morgante}, {Mortlock},
  {Moss}, {Munshi}, {Murphy}, {Naselsky}, {Nati}, {Natoli}, {Netterfield}, {N{\o}rgaard-Nielsen}, {Noviello}, {Novikov}, {Novikov}, {Oxborrow}, {Paci}, {Pagano}, {Pajot}, {Paladini}, {Paoletti}, {Partridge}, {Pasian}, {Patanchon}, {Pearson}, {Perdereau}, {Perotto}, {Perrotta}, {Pettorino}, {Piacentini}, {Piat}, {Pierpaoli}, {Pietrobon}, {Plaszczynski}, {Pointecouteau}, {Polenta}, {Popa}, {Pratt}, {Pr{\'e}zeau}, {Prunet}, {Puget}, {Rachen}, {Reach}, {Rebolo}, {Reinecke}, {Remazeilles}, {Renault}, {Renzi}, {Ristorcelli}, {Rocha}, {Rosset}, {Rossetti}, {Roudier}, {Rouill{\'e} d'Orfeuil}, {Rowan-Robinson}, {Rubi{\~n}o-Mart{\'\i}n}, {Rusholme}, {Said}, {Salvatelli}, {Salvati}, {Sandri}, {Santos}, {Savelainen}, {Savini}, {Scott}, {Seiffert}, {Serra}, {Shellard}, {Spencer}, {Spinelli}, {Stolyarov}, {Stompor}, {Sudiwala}, {Sunyaev}, {Sutton}, {Suur-Uski}, {Sygnet}, {Tauber}, {Terenzi}, {Toffolatti}, {Tomasi}, {Tristram}, {Trombetti}, {Tucci}, {Tuovinen}, {T{\"u}rler}, {Umana}, {Valenziano}, {Valiviita}, {Van Tent},
  {Vielva}, {Villa}, {Wade}, {Wandelt}, {Wehus}, {White}, {White}, {Wilkinson}, {Yvon}, {Zacchei}, \& {Zonca}}]{planck_collab2016}
{Planck Collaboration}, {Ade}, P.~A.~R., {Aghanim}, N., {et~al.} 2016, \aap, 594, A13, \dodoi{10.1051/0004-6361/201525830}

\bibitem[{Prandini \& Ghisellini(2022)}]{prandini2022}
Prandini, E., \& Ghisellini, G. 2022, Galaxies, 10, \dodoi{10.3390/galaxies10010035}

\bibitem[{{Richards} {et~al.}(2006){Richards}, {Strauss}, {Fan}, {Hall}, {Jester}, {Schneider}, {Vanden Berk}, {Stoughton}, {Anderson}, {Brunner}, {Gray}, {Gunn}, {Ivezi{\'c}}, {Kirkland}, {Knapp}, {Loveday}, {Meiksin}, {Pope}, {Szalay}, {Thakar}, {Yanny}, {York}, {Barentine}, {Brewington}, {Brinkmann}, {Fukugita}, {Harvanek}, {Kent}, {Kleinman}, {Krzesi{\'n}ski}, {Long}, {Lupton}, {Nash}, {Neilsen}, {Nitta}, {Schlegel}, \& {Snedden}}]{richards2006}
{Richards}, G.~T., {Strauss}, M.~A., {Fan}, X., {et~al.} 2006, \aj, 131, 2766, \dodoi{10.1086/503559}

\bibitem[{{Richards} {et~al.}(2009){Richards}, {Myers}, {Gray}, {Riegel}, {Nichol}, {Brunner}, {Szalay}, {Schneider}, \& {Anderson}}]{richards2009}
{Richards}, G.~T., {Myers}, A.~D., {Gray}, A.~G., {et~al.} 2009, \apjs, 180, 67, \dodoi{10.1088/0067-0049/180/1/67}

\bibitem[{{Sahakyan} {et~al.}(2020){Sahakyan}, {Israyelyan}, {Harutyunyan}, {Khachatryan}, \& {Gasparyan}}]{sahakyan2020}
{Sahakyan}, N., {Israyelyan}, D., {Harutyunyan}, G., {Khachatryan}, M., \& {Gasparyan}, S. 2020, \mnras, 498, 2594, \dodoi{10.1093/mnras/staa2477}

\bibitem[{{Sbarrato} {et~al.}(2015){Sbarrato}, {Ghisellini}, {Tagliaferri}, {Foschini}, {Nardini}, {Tavecchio}, \& {Gehrels}}]{sbarrato2015}
{Sbarrato}, T., {Ghisellini}, G., {Tagliaferri}, G., {et~al.} 2015, \mnras, 446, 2483, \dodoi{10.1093/mnras/stu2269}

\bibitem[{{Sbarrato} {et~al.}(2022){Sbarrato}, {Ghisellini}, {Tagliaferri}, {Tavecchio}, {Ghirlanda}, \& {Costamante}}]{sbarrato2022}
---. 2022, \aap, 663, A147, \dodoi{10.1051/0004-6361/202243569}

\bibitem[{{Schmidt} {et~al.}(1992){Schmidt}, {Elston}, \& {Lupie}}]{Schmidt1992}
{Schmidt}, G.~D., {Elston}, R., \& {Lupie}, O.~L. 1992, \aj, 104, 1563, \dodoi{10.1086/116341}

\bibitem[{{Schneider} {et~al.}(2007){Schneider}, {Hall}, {Richards}, {Strauss}, {Vanden Berk}, {Anderson}, {Brandt}, {Fan}, {Jester}, {Gray}, {Gunn}, {SubbaRao}, {Thakar}, {Stoughton}, {Szalay}, {Yanny}, {York}, {Bahcall}, {Barentine}, {Blanton}, {Brewington}, {Brinkmann}, {Brunner}, {Castander}, {Csabai}, {Frieman}, {Fukugita}, {Harvanek}, {Hogg}, {Ivezi{\'c}}, {Kent}, {Kleinman}, {Knapp}, {Kron}, {Krzesi{\'n}ski}, {Long}, {Lupton}, {Nitta}, {Pier}, {Saxe}, {Shen}, {Snedden}, {Weinberg}, \& {Wu}}]{schneider2007}
{Schneider}, D.~P., {Hall}, P.~B., {Richards}, G.~T., {et~al.} 2007, \aj, 134, 102, \dodoi{10.1086/518474}

\bibitem[{{Sexton} {et~al.}(2022){Sexton}, {Secrest}, {Johnson}, \& {Dorland}}]{sexton2022}
{Sexton}, R.~O., {Secrest}, N.~J., {Johnson}, M.~C., \& {Dorland}, B.~N. 2022, \apjs, 260, 33, \dodoi{10.3847/1538-4365/ac609f}

\bibitem[{{Shakura} \& {Sunyaev}(1973)}]{Shakura1973}
{Shakura}, N.~I., \& {Sunyaev}, R.~A. 1973, \aap, 24, 337

\bibitem[{{Shen} {et~al.}(2011){Shen}, {Richards}, {Strauss}, {Hall}, {Schneider}, {Snedden}, {Bizyaev}, {Brewington}, {Malanushenko}, {Malanushenko}, {Oravetz}, {Pan}, \& {Simmons}}]{shen2011}
{Shen}, Y., {Richards}, G.~T., {Strauss}, M.~A., {et~al.} 2011, \apjs, 194, 45, \dodoi{10.1088/0067-0049/194/2/45}

\bibitem[{{Shimwell} {et~al.}(2022){Shimwell}, {Hardcastle}, {Tasse}, {Best}, {R{\"o}ttgering}, {Williams}, {Botteon}, {Drabent}, {Mechev}, {Shulevski}, {van Weeren}, {Bester}, {Br{\"u}ggen}, {Brunetti}, {Callingham}, {Chy{\.z}y}, {Conway}, {Dijkema}, {Duncan}, {de Gasperin}, {Hale}, {Haverkorn}, {Hugo}, {Jackson}, {Mevius}, {Miley}, {Morabito}, {Morganti}, {Offringa}, {Oonk}, {Rafferty}, {Sabater}, {Smith}, {Schwarz}, {Smirnov}, {O'Sullivan}, {Vedantham}, {White}, {Albert}, {Alegre}, {Asabere}, {Bacon}, {Bonafede}, {Bonnassieux}, {Brienza}, {Bilicki}, {Bonato}, {Calistro Rivera}, {Cassano}, {Cochrane}, {Croston}, {Cuciti}, {Dallacasa}, {Danezi}, {Dettmar}, {Di Gennaro}, {Edler}, {En{\ss}lin}, {Emig}, {Franzen}, {Garc{\'\i}a-Vergara}, {Grange}, {G{\"u}rkan}, {Hajduk}, {Heald}, {Heesen}, {Hoang}, {Hoeft}, {Horellou}, {Iacobelli}, {Jamrozy}, {Jeli{\'c}}, {Kondapally}, {Kukreti}, {Kunert-Bajraszewska}, {Magliocchetti}, {Mahatma}, {Ma{\l}ek}, {Mandal}, {Massaro}, {Meyer-Zhao}, {Mingo}, {Mostert}, {Nair},
  {Nakoneczny}, {Nikiel-Wroczy{\'n}ski}, {Orr{\'u}}, {Pajdosz-{\'S}mierciak}, {Pasini}, {Prandoni}, {van Piggelen}, {Rajpurohit}, {Retana-Montenegro}, {Riseley}, {Rowlinson}, {Saxena}, {Schrijvers}, {Sweijen}, {Siewert}, {Timmerman}, {Vaccari}, {Vink}, {West}, {Wo{\l}owska}, {Zhang}, \& {Zheng}}]{lofar_lotts}
{Shimwell}, T.~W., {Hardcastle}, M.~J., {Tasse}, C., {et~al.} 2022, \aap, 659, A1, \dodoi{10.1051/0004-6361/202142484}

\bibitem[{{S{\k{a}}dowski} {et~al.}(2014){S{\k{a}}dowski}, {Narayan}, {McKinney}, \& {Tchekhovskoy}}]{sadowski2014}
{S{\k{a}}dowski}, A., {Narayan}, R., {McKinney}, J.~C., \& {Tchekhovskoy}, A. 2014, \mnras, 439, 503, \dodoi{10.1093/mnras/stt2479}

\bibitem[{{Smith}(2016)}]{Smith2016}
{Smith}, P.~S. 2016, Galaxies, 4, 27, \dodoi{10.3390/galaxies4030027}

\bibitem[{{Smith} {et~al.}(1991){Smith}, {Jannuzi}, \& {Elston}}]{Smith1991}
{Smith}, P.~S., {Jannuzi}, B.~T., \& {Elston}, R. 1991, \apjs, 77, 67, \dodoi{10.1086/191598}

\bibitem[{{Smith} {et~al.}(2007){Smith}, {Williams}, {Schmidt}, {Diamond-Stanic}, \& {Means}}]{Smith2007}
{Smith}, P.~S., {Williams}, G.~G., {Schmidt}, G.~D., {Diamond-Stanic}, A.~M., \& {Means}, D.~L. 2007, \apj, 663, 118, \dodoi{10.1086/517992}

\bibitem[{{Tomsick} {et~al.}(2019){Tomsick}, {Zoglauer}, {Sleator}, {Lazar}, {Beechert}, {Boggs}, {Roberts}, {Siegert}, {Lowell}, {Wulf}, {Grove}, {Phlips}, {Brandt}, {Smale}, {Kierans}, {Burns}, {Hartmann}, {Leising}, {Ajello}, {Fryer}, {Amman}, {Chang}, {Jean}, \& {von Ballmoos}}]{tomsick2019}
{Tomsick}, J., {Zoglauer}, A., {Sleator}, C., {et~al.} 2019, in Bulletin of the American Astronomical Society, Vol.~51, 98, \dodoi{10.48550/arXiv.1908.04334}

\bibitem[{{Tomsick} {et~al.}(2024){Tomsick}, {Boggs}, {Zoglauer}, {Hartmann}, {Ajello}, {Burns}, {Fryer}, {Karwin}, {Kierans}, {Lowell}, {Malzac}, {Roberts}, {Saint-Hilaire}, {Shih}, {Siegert}, {Sleator}, {Takahashi}, {Tavecchio}, {Wulf}, {Beechert}, {Gulick}, {Joens}, {Lazar}, {Neights}, {Martinez Oliveros}, {Matsumoto}, {Melia}, {Yoneda}, {Amman}, {Bal}, {von Ballmoos}, {Bates}, {B{\"o}ttcher}, {Bulgarelli}, {Cavazzuti}, {Chang}, {Chen}, {Chu}, {Ciabattoni}, {Costamante}, {Dreyer}, {Fioretti}, {Fenu}, {Gallego}, {Ghirlanda}, {Grove}, {Huang}, {Jean}, {Khatiya}, {Kn{\"o}dlseder}, {Kraus}, {Leising}, {Lewis}, {Lommler}, {Marcotulli}, {Martinez Castellanos}, {Mittal}, {Negro}, {Al Nussirat}, {Nakazawa}, {Oberlack}, {Palmore}, {Panebianco}, {Parmiggiani}, {Pike}, {Rogers}, {Schutte}, {Sheng}, {Smale}, {Smith}, {Trigg}, {Venters}, {Watanabe}, \& {Zhang}}]{tomsick2023}
{Tomsick}, J., {Boggs}, S., {Zoglauer}, A., {et~al.} 2024, in 38th International Cosmic Ray Conference, 745, \dodoi{10.48550/arXiv.2308.12362}

\bibitem[{{Truemper}(1982)}]{truemper1982}
{Truemper}, J. 1982, Advances in Space Research, 2, 241, \dodoi{10.1016/0273-1177(82)90070-9}

\bibitem[{{Urry} \& {Padovani}(1995)}]{urry1995}
{Urry}, C.~M., \& {Padovani}, P. 1995, \pasp, 107, 803, \dodoi{10.1086/133630}

\bibitem[{{Veres} {et~al.}(2010){Veres}, {Frey}, {Paragi}, \& {Gurvits}}]{veres2010}
{Veres}, P., {Frey}, S., {Paragi}, Z., \& {Gurvits}, L.~I. 2010, \aap, 521, A6, \dodoi{10.1051/0004-6361/201014957}

\bibitem[{{Verner} {et~al.}(1996){Verner}, {Ferland}, {Korista}, \& {Yakovlev}}]{vern}
{Verner}, D.~A., {Ferland}, G.~J., {Korista}, K.~T., \& {Yakovlev}, D.~G. 1996, \apj, 465, 487, \dodoi{10.1086/177435}

\bibitem[{{Volonteri} {et~al.}(2011){Volonteri}, {Haardt}, {Ghisellini}, \& {Della Ceca}}]{volonteri2011}
{Volonteri}, M., {Haardt}, F., {Ghisellini}, G., \& {Della Ceca}, R. 2011, \mnras, 416, 216, \dodoi{10.1111/j.1365-2966.2011.19024.x}

\bibitem[{{Volonteri} {et~al.}(2003){Volonteri}, {Haardt}, \& {Madau}}]{Volonteri2003}
{Volonteri}, M., {Haardt}, F., \& {Madau}, P. 2003, \apj, 582, 559, \dodoi{10.1086/344675}

\bibitem[{{Wang} {et~al.}(2021){Wang}, {Yang}, {Fan}, {Hennawi}, {Barth}, {Banados}, {Bian}, {Boutsia}, {Connor}, {Davies}, {Decarli}, {Eilers}, {Farina}, {Green}, {Jiang}, {Li}, {Mazzucchelli}, {Nanni}, {Schindler}, {Venemans}, {Walter}, {Wu}, \& {Yue}}]{wang2021}
{Wang}, F., {Yang}, J., {Fan}, X., {et~al.} 2021, \apjl, 907, L1, \dodoi{10.3847/2041-8213/abd8c6}

\bibitem[{{Watson} {et~al.}(2009){Watson}, {Schr{\"o}der}, {Fyfe}, {Page}, {Lamer}, {Mateos}, {Pye}, {Sakano}, {Rosen}, {Ballet}, {Barcons}, {Barret}, {Boller}, {Brunner}, {Brusa}, {Caccianiga}, {Carrera}, {Ceballos}, {Della Ceca}, {Denby}, {Denkinson}, {Dupuy}, {Farrell}, {Fraschetti}, {Freyberg}, {Guillout}, {Hambaryan}, {Maccacaro}, {Mathiesen}, {McMahon}, {Michel}, {Motch}, {Osborne}, {Page}, {Pakull}, {Pietsch}, {Saxton}, {Schwope}, {Severgnini}, {Simpson}, {Sironi}, {Stewart}, {Stewart}, {Stobbart}, {Tedds}, {Warwick}, {Webb}, {West}, {Worrall}, \& {Yuan}}]{xmm_serendipitous_source_catalog}
{Watson}, M.~G., {Schr{\"o}der}, A.~C., {Fyfe}, D., {et~al.} 2009, \aap, 493, 339, \dodoi{10.1051/0004-6361:200810534}

\bibitem[{{Whalen} \& {Fryer}(2012)}]{whalen2012}
{Whalen}, D.~J., \& {Fryer}, C.~L. 2012, \apjl, 756, L19, \dodoi{10.1088/2041-8205/756/1/L19}

\bibitem[{{Wilms} {et~al.}(2000){Wilms}, {Allen}, \& {McCray}}]{wilms_tbabs}
{Wilms}, J., {Allen}, A., \& {McCray}, R. 2000, \apj, 542, 914, \dodoi{10.1086/317016}

\bibitem[{{Wise} {et~al.}(2019){Wise}, {Regan}, {O'Shea}, {Norman}, {Downes}, \& {Xu}}]{wise2019}
{Wise}, J.~H., {Regan}, J.~A., {O'Shea}, B.~W., {et~al.} 2019, \nat, 566, 85, \dodoi{10.1038/s41586-019-0873-4}

\bibitem[{{Wolf} {et~al.}(2024){Wolf}, {Salvato}, {Belladitta}, {Arcodia}, {Ciroi}, {Di Mille}, {Sbarrato}, {Buchner}, {H{\"a}mmerich}, {Wilms}, {Collmar}, {Dwelly}, {Merloni}, {Urrutia}, \& {Nandra}}]{wolf2024}
{Wolf}, J., {Salvato}, M., {Belladitta}, S., {et~al.} 2024, \aap, 691, A30, \dodoi{10.1051/0004-6361/202451035}

\bibitem[{{Wood} {et~al.}(2017){Wood}, {Caputo}, {Charles}, {Di Mauro}, {Magill}, {Perkins}, \& {Fermi-LAT Collaboration}}]{fermipy}
{Wood}, M., {Caputo}, R., {Charles}, E., {et~al.} 2017, in International Cosmic Ray Conference, Vol. 301, 35th International Cosmic Ray Conference (ICRC2017), 824, \dodoi{10.22323/1.301.0824}

\bibitem[{{Worsley} {et~al.}(2004){Worsley}, {Fabian}, {Celotti}, \& {Iwasawa}}]{Worsley2004}
{Worsley}, M.~A., {Fabian}, A.~C., {Celotti}, A., \& {Iwasawa}, K. 2004, \mnras, 350, L67, \dodoi{10.1111/j.1365-2966.2004.07887.x}

\bibitem[{{Worsley} {et~al.}(2006){Worsley}, {Fabian}, {Pooley}, \& {Chandler}}]{worsley2006}
{Worsley}, M.~A., {Fabian}, A.~C., {Pooley}, G.~G., \& {Chandler}, C.~J. 2006, \mnras, 368, 844, \dodoi{10.1111/j.1365-2966.2006.10173.x}

\bibitem[{{Yu} \& {Tremaine}(2002)}]{Yu2002}
{Yu}, Q., \& {Tremaine}, S. 2002, \mnras, 335, 965, \dodoi{10.1046/j.1365-8711.2002.05532.x}

\bibitem[{{Zhang} {et~al.}(2020){Zhang}, {An}, \& {Frey}}]{zhang2020}
{Zhang}, Y., {An}, T., \& {Frey}, S. 2020, Science Bulletin, 65, 525, \dodoi{10.1016/j.scib.2020.01.008}

\bibitem[{{Zhou} {et~al.}(2021){Zhou}, {Dai}, \& {Yang}}]{zhou2021}
{Zhou}, B., {Dai}, B., \& {Yang}, J. 2021, \pasj, 73, 850, \dodoi{10.1093/pasj/psab051}

\bibitem[{{Zombeck} {et~al.}(1995){Zombeck}, {David}, {Harnden}, \& {Kearns}}]{zombeck1995}
{Zombeck}, M.~V., {David}, L.~P., {Harnden}, F.~R., \& {Kearns}, K. 1995, in Society of Photo-Optical Instrumentation Engineers (SPIE) Conference Series, Vol. 2518, EUV, X-Ray, and Gamma-Ray Instrumentation for Astronomy VI, ed. O.~H. {Siegmund} \& J.~V. {Vallerga}, 304--321, \dodoi{10.1117/12.218385}

\end{thebibliography}
\bibliographystyle{aasjournal}



\end{document}